%% file: ms.tex
\documentclass[numberedappendix]{emulateapj}

\usepackage{graphicx}
\usepackage{natbib}
\usepackage{apjfonts}
\usepackage{amsmath}
\usepackage{color}
\usepackage{comment}
\usepackage{rotating}
\usepackage{url}
\bibpunct{(}{)}{;}{a}{}{,}
\usepackage{lineno}
\usepackage{gensymb}

\newcommand{\finalcut}{\textsc{Finalcut}~}
\newcommand{\sextractor}{\textsc{SExtractor}~}
\newcommand{\swarp}{\textsc{SWARP}~}
\newcommand{\rdmp}{\textsc{redMaPPer}~}
\newcommand{\ngmix}{\textsc{ngmix}~}

\begin{document}
\begin{nolinenumbers}
\vspace*{-\headsep}\vspace*{\headheight}
\footnotesize \hfill FERMILAB-PUB-18-663-AE-E\\
\vspace*{-\headsep}\vspace*{\headheight}
\footnotesize \hfill DES-2017-0244
\end{nolinenumbers}

\title{Dark Energy Survey Year 1 results: Detection of Intracluster Light at Redshift $\sim$ 0.25}

\input{icl_authors.tex}
\email{\hspace{1em}$^\dagger$E-mail: Yuanyuan Zhang, ynzhang@fnal.gov}

\begin{abstract}
Using data collected by the Dark Energy Survey (DES), we report the detection of intracluster light (ICL)  with $\sim300$ galaxy clusters in the redshift range of 0.2-0.3. We design methods to mask detected galaxies and stars in the images and stack the cluster light profiles, while accounting for several systematic effects (sky subtraction, instrumental point-spread function, cluster selection effects and residual light in the ICL raw detection from background and cluster galaxies). The methods allow us to acquire high signal-to-noise measurements of the ICL and central galaxies (CGs), which we separate with radial cuts. The ICL appears as faint and diffuse light extending to at least 1 Mpc from the cluster center, reaching a surface brightness level of 30 mag arcsec$^{-2}$.  The ICL and the cluster CG contribute to $44\%\pm17$\% of the total cluster stellar luminosity within 1 Mpc. The ICL color is overall consistent with that of the cluster red sequence galaxies, but displays the trend of becoming bluer with increasing radius. The ICL demonstrates an interesting self-similarity feature -- for clusters in different richness ranges,  their ICL radial profiles are similar after scaling with cluster $R_\mathrm{200m}$, and the ICL brightness appears to be a good tracer of the cluster radial mass distribution.   These analyses are based on the DES redMaPPer cluster sample identified in the first year of observations. 
\end{abstract}
\maketitle

\section{Introduction}

The central galaxy (CG) of a galaxy cluster tends to be surrounded by an extended light envelope \citep{1951PASP...63...61Z, 1952PASP...64..242Z, 1964ApJ...140...35M, 1965ApJ...142.1364M}. Studies indicate that this light envelope extends to hundreds of kiloparsecs and sometimes encloses several galaxies, especially if the cluster is experiencing a merging process \citep[see reviews in][]{2004cgpc.symp..277M, 2014ApJ...797...82L, 2015IAUGA..2247903M}. Given its diffuse nature \citep{1986ApJS...60..603S, 1987ApJS...64..643S, 1988ApJ...328..475S} and the fact that it may enclose multiple galaxies, it seems more reasonable to consider this envelope as a genuine component of galaxy clusters. The diffuse light envelope is thus frequently referred to as the intracluster light (ICL).

Despite the conceptual difference between CGs and the ICL,  it is almost impossible to observationally separate them, because the outskirts of CGs naturally blend into the ICL. Outside the outskirts of CGs, the ICL is extremely faint, which poses significant challenges for separating it from CGs, and for further characterizing its distribution and properties. In this paper, we do not attempt to dissect the ICL and CGs. Instead, we follow a similar convention to \cite{2018MNRAS.475..648P}, wherein we simultaneously characterize CGs and the ICL. We use "ICL" to qualitatively refer to the diffuse light outside a few kpcs of the CG centers, and when making quantitative assessments, "ICL" and "CG" are quantitatively distinguished by radial cuts.

Studies, especially of the halo occupation distribution, indicate that the ICL potentially makes up a significant fraction of the cluster stellar light. Through modeling the growth histories of cluster member galaxies or cluster CGs, a few analyses have found that much of the stellar mass accreted by a galaxy cluster is missing from the luminosity or stellar mass sum of cluster galaxies or CGs, and the ICL has been considered the key to this missing mass \citep{2004ApJ...617..879L, 2006ApJ...652L..89M, 2012MNRAS.425.2058B, 2014MNRAS.437.3787C, 2007ApJ...668..826C}. According to the extrapolation in \cite{2013ApJ...770...57B},  the ICL may contain $\sim$ 2.5-5 times the stellar mass of the CG for a $1\times 10^{14}M_\odot$ cluster. Although a quantitative conclusion about ICL stellar content is yet to be reached in simulations and observational studies,  many estimate that the ICL and CGs consist of up to $\sim$ 10 \% to 50 \% of the total cluster stellar content \citep[e.g., ][]{2005MNRAS.358..949Z, 2007ApJ...666..147G,  2015MNRAS.449.2353B, 2018MNRAS.474..917M}.

The origin and distribution of the ICL have been explored theoretically or through simulation studies. Most of these explain the origin of the ICL as stars that are originally formed inside galaxies and became dispersed into the intracluster space during galaxy interactions \citep[see reviews in ][]{2014MNRAS.437.3787C, 2018MNRAS.474.3009D}. Tidal stripping \citep{1972AJ.....77..288G}, galaxy disruption \citep{2011MNRAS.413..101G} and relaxation of galaxy mergers \citep{2007MNRAS.377....2M} are possible astrophysical processes that produce the ICL in this way. Surprisingly, in-situ star formation also appears to be a viable contributor to the ICL \citep{2010MNRAS.406..936P, doi:10.1111/j.1365-2966.2012.20737.x}. The ICL likely has formed through multiple channels, but the relative contributions from the channels are still being explored. Depending on the main formation mechanism, simulated ICL exhibits different color and spatial distributions \citep{2018arXiv181103253C}, and the total amount of the ICL stellar mass varies \citep{2018MNRAS.479..932C}, which provide clues for testing ICL formation hypotheses with observations. 

Observational studies of the ICL have been performed with both {\it Hubble} and ground-based telescopes. These include targeted imaging observations of individual \citep[e.g., ][]{2006AJ....131..168K, 2011MNRAS.414..602T, 2014ApJ...781...24G, 2014ApJ...794..137M, 2014A&A...565A.126P}
 or a sample of clusters \citep[e.g., ][]{2007AJ....134..466K, 2017ApJ...846..139M, 2018MNRAS.474.3009D, 2018MNRAS.474..917M}, or statistical studies of  a few hundreds of clusters based on wide field survey data \citep{2005MNRAS.358..949Z}. Many works focused on nearby \citep[e.g.,][]{2017ApJ...834...16M} or intermediate-redshift clusters, but high-redshift ICL observations to $z\sim 1.0$ \citep{2012MNRAS.425.2058B, 2015MNRAS.449.2353B, 2018ApJ...862...95K} have been reported. Advances in integral-field spectroscopy also bring forward spectroscopic studies of the ICL stellar population \citep{2016A&A...592A...7A, 2016MNRAS.461..230E, 2018arXiv180504520G, 2018arXiv180506913J}. The ICL has been found to contain an old stellar population and extends to several hundreds of parsecs from cluster centers, although cluster-to-cluster variation \citep[e.g.,][]{2007AJ....134..466K}, cluster dynamic state \citep{2018ApJ...857...79J}, redshift evolution in addition to methodological differences and sometimes differences in ICL definitions \citep{2014MNRAS.437..816C, 2016ApJ...820...49J, 2018arXiv180403335T} may cause  different conclusions.
 
Due to its low surface brightness and diffuseness, studying the ICL presents significant challenges. Simulation ICL studies generally require high resolution \citep{2014MNRAS.437.3787C} as well as proper baryonic physics such as active galactic nucleus (AGN) feedback \citep{ 2016MNRAS.459.4408M, 2018MNRAS.475..648P} to trace the dispersion of stellar particles. Observational studies using the {\it Hubble Space Telescope} (HST) tend to be pointed observations and benefit from the high resolution and low sky background level of the images but are limited to small cluster samples. ICL studies with ground-based telescopes benefit from the richer observing resources yet face more challenges in data processing. Studies of individual galaxy clusters require long exposure time, often dozens of hours even with large-aperture telescopes. Analyses of the data then require a meticulous account of systematic effects such as flat-fielding and sky background subtraction, and the telescope and imaging camera point spread function (PSF) is a confusing effect for interpreting ICL detection results at large cluster radius \citep{2007ApJ...666..663B, 2014A&A...567A..97S, 2018arXiv180403335T}.

Studying the statistical distribution of the ICL in an ensemble of clusters is one approach toward easing systematic effects from ground-based observations. A particularly exciting method is to combine the images of dozens to several hundreds of galaxy clusters by aligning the centers of the CGs and measuring the ICL in these combined images, also known as ``stacking''. With this approach, \citet{2005MNRAS.358..949Z} analyzed $\sim 600$ clusters with Sloan Digital Sky Survey (SDSS) data and detected the ICL at a surface brightness level of 32 mag/arcsec$^{2}$. \cite{2006AJ....131..168K} adopted a similar ICL analysis method, and \cite{2011ApJ...731...89T} and \cite{2018arXiv181104714W} applied the stacking idea to analyze the light envelopes of luminous red galaxies and the CGs of galaxy groups. Notably, the stacking method is less affected by, but not completely immune to, systematic effects. Issues such as sky background subtraction and the PSF still plague the  method and need to be carefully evaluated.

In this paper, we use a ``stacking'', i.e., an averaging approach in which we analyze hundreds of clusters for ICL detection and characterization. We use observational data sets from the Dark Energy Survey (DES), which is in the process of collecting imaging data of an unprecedented combination of large area (5000 deg$^2$ of the sky) and depth. DES expects to find 30,000 galaxy clusters using their population of member galaxies. Although this paper focuses on ICL detection in clusters of redshift $0.2 - 0.3$ identified in the first year of DES observations (DES Y1), the final DES data set will eventually allow us to apply our methods to a higher redshift range and possibly characterize the redshift evolution of ICL. 

We develop a ``stacking'' approach in that we study the ICL by stacking the light profiles derived from individual cluster images, rather than  stacking the images of the clusters as in previous works \citep{2005MNRAS.358..949Z, 2006AJ....131..168K, 2011ApJ...731...89T}. Our method achieves results equivalent to the latter method upon our initial method testing, but it is computationally more efficient and versatile. The data sets and methods used in this paper are described in Sections~\ref{sec:data} and \ref{sec:methods} respectively.
Because PSF is a possible effect that influences ICL interpretation, we derive the DES Dark Energy Camera PSFs in Section~\ref{sec:psf}. Section~\ref{sec:ICL} and Section~\ref{sec:clusters} present our results on ICL detection and ICL properties. In Section~\ref{sec:systematics}, we discuss how systematic effects, such as sky background subtraction and the PSF, influence the interpretation of the results. Section~\ref{sec:conclude} discusses and summarizes the analyses. Throughout the paper, we assume a flat $\Lambda$CDM cosmology with $h=0.7$.

\section{Data}
\label{sec:data}

\subsection{The \rdmp Cluster Catalogs}
\label{sec:redmapper_clusters}

\begin{figure}
\includegraphics[width=0.55\textwidth]{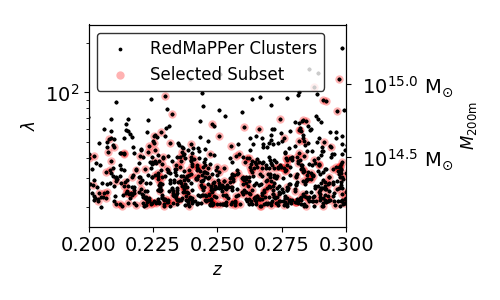}
\caption{Redshifts and richnesses of the \rdmp clusters used in this work. The masses of the clusters are estimated from richnesses using the mass-richness relation in Melchior et al. 2017. The mean mass of the clusters studied in this paper, which pass the purging criteria in Section~\ref{sec:redmapper_clusters} (red shaded symbols, the "Selected Subset"), is $2.50\times10^{14}\mathrm{M_\odot}$.}
\label{fig:rdmp_z}
\end{figure}

We study the ICL in a galaxy cluster sample identified with the \rdmp algorithm \citep{2014ApJ...785..104R}. \rdmp searches for galaxy clusters by considering galaxy colors, luminosities, and spatial distributions. A notable feature of the algorithm is that it looks for cluster galaxies that match the colors of the red sequence galaxies in spectroscopically confirmed galaxy clusters. The result is a highly complete and pure cluster sample with precise photometric redshifts and a cluster richness estimator that scales well with cluster mass \citep{2012ApJ...746..178R, 2014ApJ...783...80R, 2015MNRAS.450..592R, 2015MNRAS.453...38R, 2017MNRAS.469.4899M, 2018arXiv180500039M}. A detailed description of the algorithm can be found in \cite{2016ApJS..224....1R}. The final products of the \rdmp algorithm includes a cluster catalog, a cluster member galaxy catalog, and a random point catalog that records the sky coverage of the cluster-finding algorithm. Both the cluster catalog and the random point catalog are used in this paper.

Specifically, we select a subset of  $0.2<z<0.3$  \rdmp clusters discovered in DES Year 1 coadd observations \citep{2018arXiv180500039M}. Because of a slightly better performance in identifying cluster CGs, we use a DES internal version of the \rdmp sample based on DESDM coadd catalogs, rather than the nominal \rdmp sample based on multiobject fitting photometry \citep{2017arXiv170801531D}. We eliminate galaxy clusters that have flags of bright stars, bright galaxies, adjacency to the Large Magellanic Cloud, etc. \citep[bad region mask $>$ 2 in][]{2017arXiv170801531D} within 526\arcsec (2000 DES coadd image pixels) of the \rdmp centers. In particular, these eliminations reduce undesirable imaging features such as streaks of saturated stars or improper sky estimation around nearby galaxies, which would lower the overall signal-to-noise ratio (S/N) of ICL detection. A DES object depth cut is also applied, with details described in Section ~\ref{sec:des_coadd_obj}, which requires object detection to be highly complete in the cluster regions. These criteria remove $\sim$ 2/3 of the \rdmp cluster sample. In total, we analyze $\sim$ 280 galaxy clusters with a mean mass of $2.50\times10^{14}\mathrm{M_\odot}$, the redshift and richness distribution of which are shown in Figure~\ref{fig:rdmp_z}.  

We use the \rdmp random point catalog, which is a sample of random locations that traces the sky coverage of the \rdmp cluster sample and matches the sky coordinates of the \rdmp cluster search area. The main purpose of using it is to estimate light profiles of random fields in the \rdmp footprint. As described in Section~\ref{sec:methods}, the light profiles of the random fields are subtracted from the cluster light profiles to eliminate sky-subtraction residual. For this purpose, we draw a sample of 2000 random points, which is $\sim 7$ times the size of the cluster catalog to avoid significantly increasing profile uncertainties.  We also test the ICL detection procedures with the random points as described in Section~\ref{sec:random_test}. For that purpose, we draw a second set of random points of the same size as the cluster sample. 

\subsection{DES \finalcut Images}
\label{sec:finalcut}

This analysis makes use of fully processed single-exposure images, i.e., \finalcut images, from the official DES image-processing pipeline \citep{2018arXiv180103177M}. The corresponding pipeline handles bias and flat corrections, astrometry and photometry calibrations, as well as the masking of bad pixels and artifacts including cosmic rays and satellite trails. Because of the significantly improved sky-subtraction procedure applied to  DES images collected in the first three years (Y3), we make use of the \finalcut images from the DES Y3 data processing campaign, yet limited to the region of Y1 \rdmp identifications .

During this Y3 data processing campaign, the imaging sky background is estimated and subtracted from the \finalcut images using a Principal Component Analysis (PCA) method developed in \citet{Bernstein:2017}. The method evaluates the sky level across the whole focal plane image of 3 deg$^2$. The full exposure sky background is described as the linear combination of a small set (three or four) of fixed patterns, and the coefficient for each is determined independently for each exposure. This PCA method is designed to capture the possible realizations of signals produced by the uniform zodiac and atmospheric background with gentle gradients across the image, but it does not attempt to remove light coming from individual objects. Because the sky background is evaluated over the 3 deg$^2$ focal plane, it is unlikely to be severely influenced by bright stars, ghosts, dust reflection, and ICL, as long as they do not affect a significant fraction of the focal plane.  We further investigate the effect of sky subtraction on ICL detection in Section~\ref{sec:systematics}.


\subsection{The Object Catalog}
\label{sec:des_coadd_obj}

To enable the detection of ICL, we mask regions associated with objects already detected by the DES coaddition pipeline. 

The object-detection scheme from the DES coadd image is described in \cite{2017arXiv170801531D, 2018arXiv180103181A, 2018arXiv180103177M}. In short, the \sextractor software \citep{1996A&AS..117..393B} is employed to detect objects in  the DES coadd images, yielding coordinates as well as photometry measurements for the detected objects. Of particular interest to this analysis, \sextractor evaluates a Kron magnitude for each object, which is the flux enclosed within 2.5 Kron radii \citep{1980ApJS...43..305K}, denoted as \texttt{MAG\_AUTO}. The coadd catalog generated from DES Y1 data reaches a 10 $\sigma$ magnitude limit of $\sim$ 22.5 mag in the $i-$band. Although not  relevant to this paper, multiepoch, multiobject fitting photometry is also acquired for the detected objects using the \ngmix code \citep{2014MNRAS.444L..25S, 2016MNRAS.460.2245J}.

We select a subset of clusters around which object detection is highly complete and hence the objects can be correctly masked. The requirement is that the $i$-band completeness limit \citep[see definition in][]{2017arXiv171005908Z} within 526\arcsec (2000 DES coadd image pixels) of the \rdmp CGs reaches 22.4 mag. Most of the \rdmp clusters identified in Y1 data satisfy this requirement. 

\section{Methods}
\label{sec:methods}

\begin{figure*}
\includegraphics[width=1.0\textwidth]{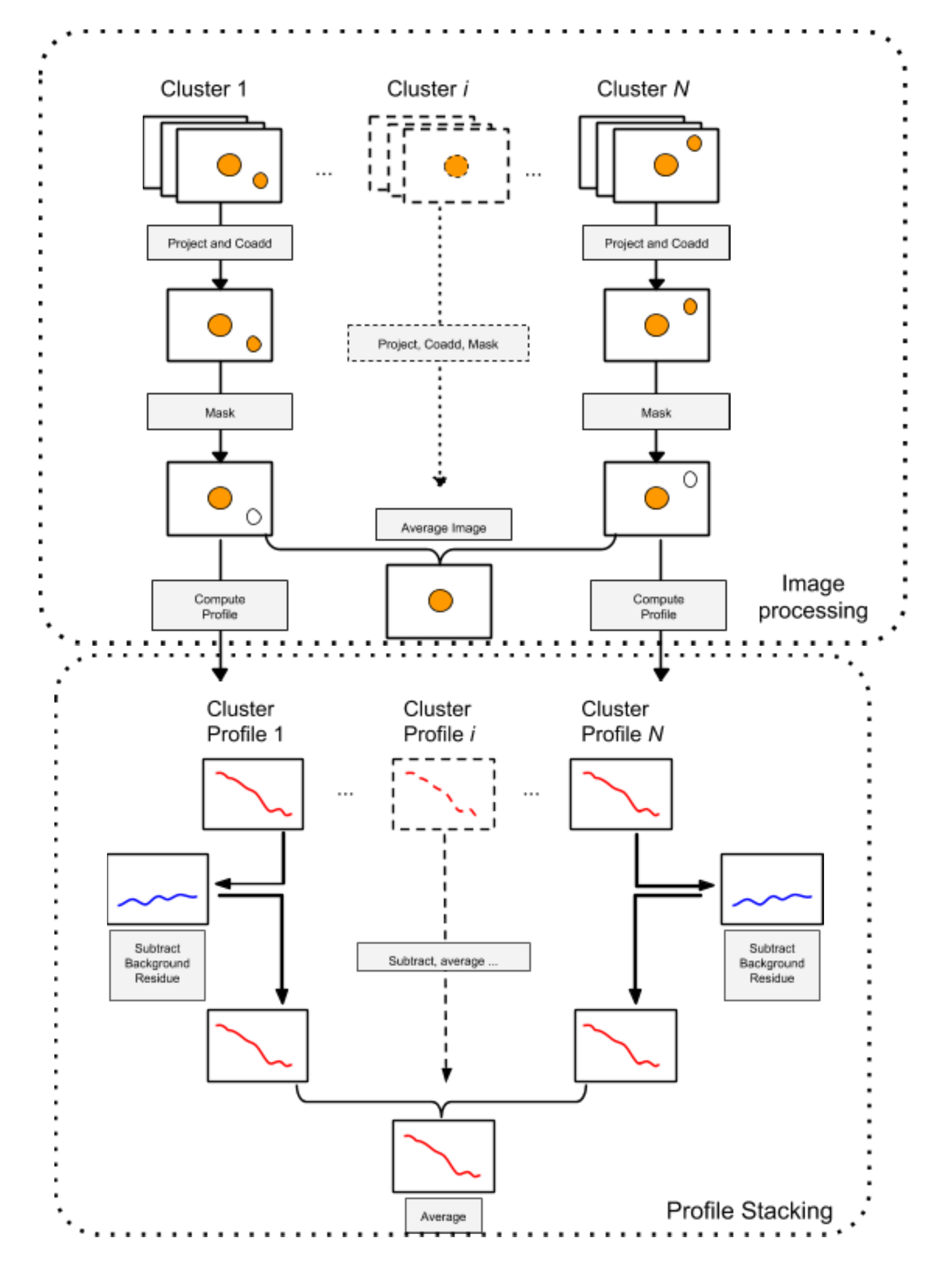}
\caption{The ICL analysis in this paper starts with DES single exposure images. The single exposure images are stacked, the detected objects are masked, and a raw ICL+CG light profile is derived for each cluster. These raw measurements go through a residual light subtraction process which removes residual background and cluster selection effects. The light profiles are finally averaged over all the clusters to create the stacked ICL+CG measurements.}
\label{fig:flowchart}
\end{figure*}

The centerpiece of this analysis is the measurement of ICL profiles. The derivation of ICL light profiles is based on DES images and consists of two steps, image processing and profile estimation. In this section, we provide the details of the method components and present a test to demonstrate the effectiveness of the methods in determining an accurate ICL light profile. The general flow of the methods is illustrated in Figure~\ref{fig:flowchart}. 

\subsection{Image Processing}

The ICL analysis starts with processing DES single-exposure images. For each of the clusters, we download DES \finalcut images and the corresponding weight maps overlapping a circular region of 0\degree.15 radius centered on the \rdmp central galaxy. These images are projected and averaged with weights to create one coadd image for each cluster, using the \swarp software \citep{2002ASPC..281..228B}. Specifically, the images are projected with a TAN projection system \citep[see a review in][]{2002A&A...395.1077C}, and the \swarp sky background function is turned off. These coadded images for each of the clusters are the input to computing the profiles of ICL. Notably, we do not use the coadded images from the official DES coaddition pipeline, which has incorporated a sky-subtraction process by the \sextractor software and thus complicates ICL detection.

To improve the detection of faint diffuse light, we mask detected objects above a magnitude limit. These objects are selected from the DES coadd catalog (Section~\ref{sec:des_coadd_obj}) to be (1) above the 5 $\sigma$ magnitude limit in the $i$-band, i.e., $\texttt{MAGERR\_AUTO\_I} < 0.218 $, and (2) i.e., brighter than 22.4 mag in the $i-$band, i.e., $\texttt{MAG\_AUTO\_I} < 22.4 $. According to the study of the relation between DES catalog depth and completeness in \cite{2017arXiv171005908Z}, objects of $i-$band $\texttt{MAG\_AUTO} < 22.4$ mag and $magerr\_auto < 0.218$ (flux measurement at 5$\sigma$) reach a $\sim$ 99.8\% completeness. We expect the depth and completeness relation in DES Y1 data to be similar to those in \cite{2017arXiv171005908Z}  because of consistent DES imaging/coadding strategies over the years.

For each of the objects, we mask the region enclosed within its 2.5 Kron radius \citep{1980ApJS...43..305K} ellipse. The major and minor axes of the ellipse as well as its inclination are computed by \sextractor in the DES coadd catalog. Analytical studies indicate that more than 90\% of the galaxy luminosity is enclosed within this elliptical aperture \citep{2005PASA...22..118G}. To ensure that bright stars are properly masked, we set a minimum masking radius along the major axis of the ellipse to be $48-2\times mag\_auto\_i$ pixels (1 pixel = 0.263\arcsec) according to a visual check of the relation between Kron magnitudes and apertures.

The CG identified by the \rdmp algorithm is excluded in the masking procedure.



\subsection{Light profile estimation}
\label{method:profile}

\begin{figure}
\includegraphics[width=0.45\textwidth]{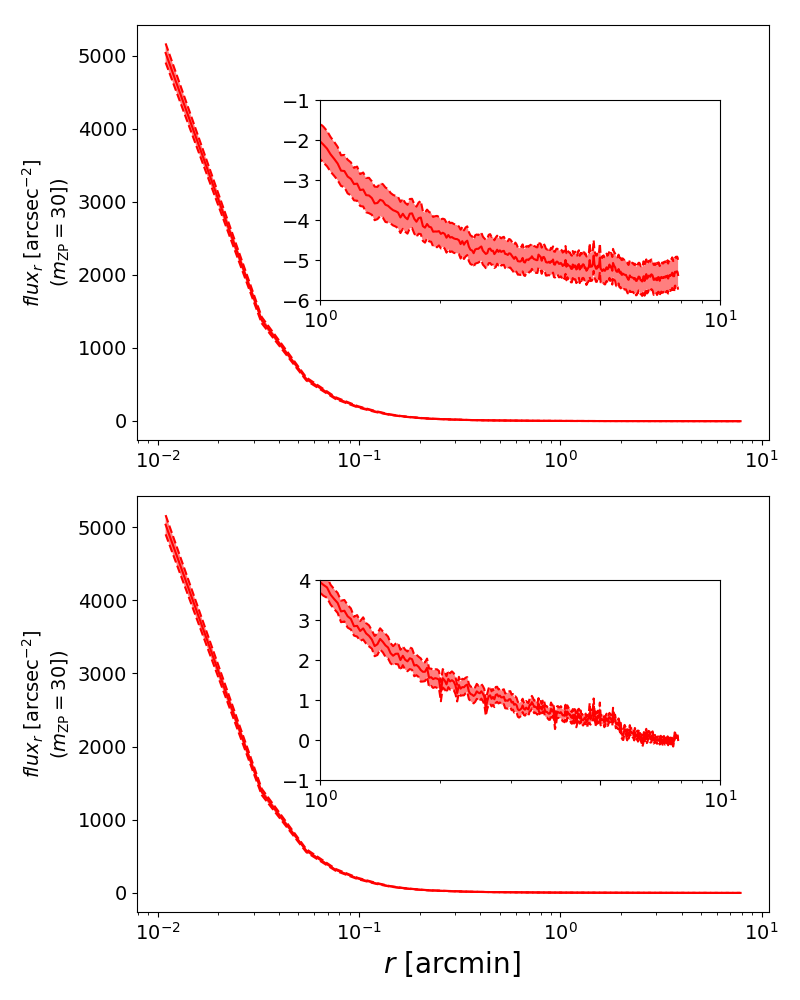}
\caption{This figure demonstrates the effect of random field subtractions on the ICL+CG profile measurement. The upper panel shows the stacked cluster light profile (after object masking) before subtracting the profiles of random fields. The lower panel shows the stacked light profile of the same clusters after subtracting random fields and residual background. The insets zoom onto the profiles at large cluster radius. The subtraction steps reduce the correlated noise and eliminate the contamination light from the cluster selection effects. The effectiveness of the subtractions is more evident in the random test shown in Figure~\ref{fig:random_process}.}
\label{fig:cluster_process}
\end{figure}

We measure the light profiles of each of the clusters from the coadded images. The light flux profiles are computed as the averaged pixel values in radial annuli centered on the \rdmp CGs, excluding the masked regions. The profiles are then averaged, i.e. "stacked", for an ensemble of clusters to detect ICL.


The cluster selection criteria eliminate undesirable features near clusters because cluster centers are selected to be 0\degree.15 away from bright stars or nearby galaxies, but the extended profiles of these undesirable features may still contribute to ICL detection, which is stronger at the cluster outskirt, causing a nonflat radial trend (see the top panels of Figures~\ref{fig:cluster_process} and~\ref{fig:random_process}). To offset this trend, we subtract the light profiles of \rdmp random points selected with the same criteria in terms of sky coverage (Section~\ref{sec:redmapper_clusters}).

We assign the clusters to 40 sky regions\footnote{This number is high enough to estimate standard deviations, while keeping each region sufficiently large to include noise caused by large-scale structures, e.g., superclusters, filaments, etc.} using a {\sc kmeans} clustering algorithm \citep{steinhaus1956division, macqueen1967} implemented for celestial coordinates\footnote{https://github.com/esheldon/kmeans\_radec}. The 2000 \rdmp random points are also assigned to the 40 regions. We measure the light profiles of each of the random points and then average them in each of the 40  {\sc kmeans} sky regions.  For each of the cluster light profiles, we subtract from it the averaged random light profile of its corresponding {\sc kmeans} region. To further reduce cluster-to-cluster variations of the background level, for each cluster, we derive the average flux level value at the edge of the image of each cluster which is 0\degree.145 from the cluster centers, and subtract it from the overall cluster light radial profile. Thus by design, for each cluster, its radial flux profile reaches 0 at 0.145 deg, which is 1.73 to 2.34 Mpc from the cluster centers depending on the cluster redshift, which is outside the $R_{200c}$ of a typical \rdmp cluster \citep{2018arXiv180500039M}.

The above region associations are also used to estimate light profile uncertainties. After the subtraction procedures, for each of the 40 aforementioned {\sc kmeans} regions, we average the cluster light profiles in the region to acquire a "stacked" measurement. The profiles of the 40 {\sc kmeans} regions are then combined using the jackknife technique \citep{lovasz1986cbms} to estimate the mean and uncertainties of the final stacked profile.

In Figure ~\ref{fig:cluster_process}, we show the final stacked profile of  the clusters before and after the subtraction procedures, estimated with the jackknife technique.  The subtraction steps eliminate sky residual and reduce correlated noise in the final stacked profile, which is more evident in the random test shown in the next section. If not explicitly stated, uncertainties presented in this paper are all estimated using the jackknife approach. Note that, by construction, they thus do not represent the cluster-to-cluster variation of the profiles.

\subsection{Test with Random Points}

\label{sec:random_test}
\begin{figure}
\includegraphics[width=0.45\textwidth]{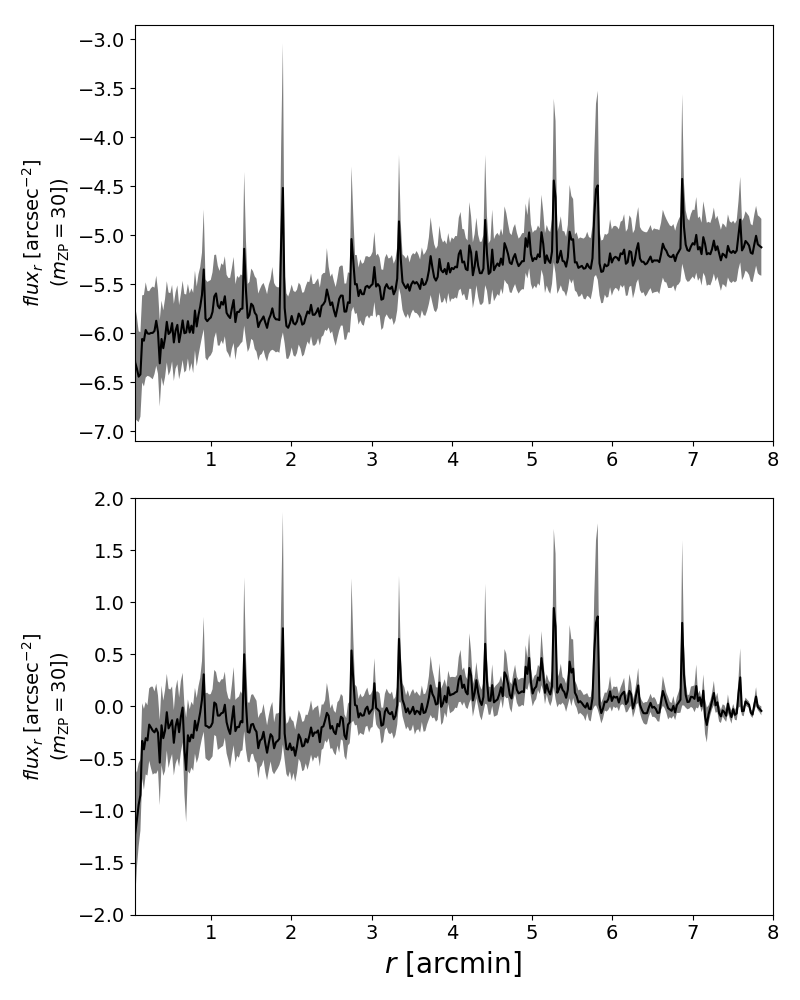}
\caption{ We apply our ICL detection methods to a set of random points of similar size to the cluster sample.  This figure (black lines) demonstrates the effectiveness of random field subtractions in deriving the stacked light profile of random points. The upper panel shows the stacked profile (after object masking) before subtracting the profiles of random fields. The lower panel shows the stacked light profile after subtracting random fields and the residual background. The subtraction steps reduce the correlated noise and eliminate the contamination light from cluster selection effects. In the lower panel, the stacked profile of the random points is consistent with 0, which is the evidence of unbiased measurement.}
\label{fig:random_process}
\end{figure}

\begin{figure}
\includegraphics[width=0.5\textwidth]{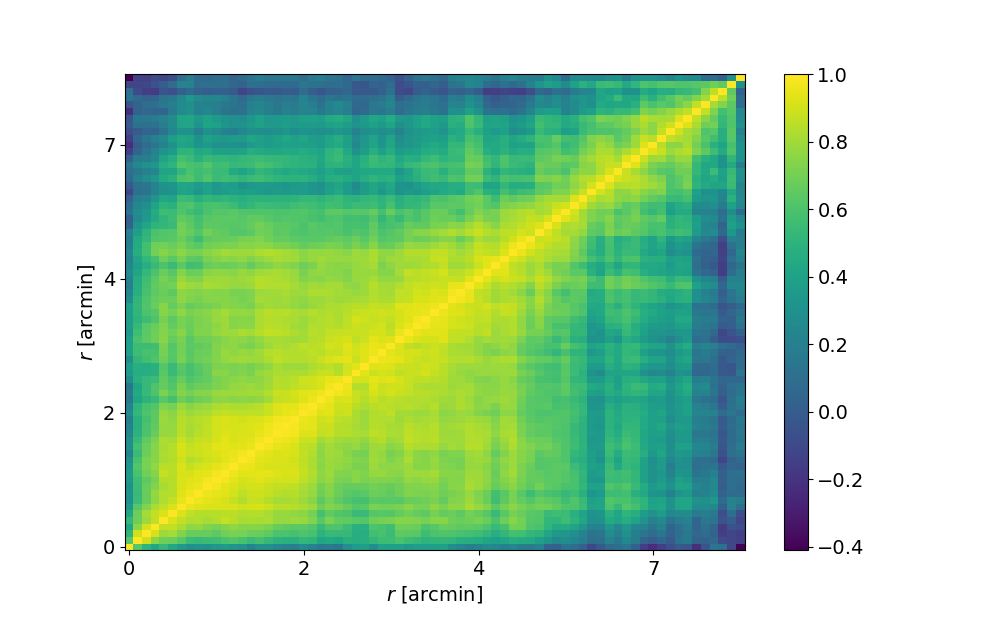}
\caption{Correlation factor of the stacked profile of the random points after the subtraction procedures. The profile appears to be correlated on a scale of a few arcseconds. This is because image imperfections are likely to affect a few pixels at a time, causing correlated  fluctuations. The Appendix lists a couple of these examples. }
\label{fig:random_profile}
\end{figure}

We first apply the methods to a set of \rdmp random points of the same size as the cluster sample. These are selected from the overall set of random points in the same way as our cluster sample from the \rdmp catalog. The set of random points goes through the image- and profile-processing steps described in earlier sections. Figure~\ref{fig:random_process}  shows the stacked light profiles of this random point sample, before and after the profile-subtraction procedures (Section~\ref{method:profile}). The stacked light profiles before the subtraction procedures show a nonflat trend from the center of the image to the outskirt. After image inspections and noticing that random points/clusters are at a distance away from bright stars or nearby galaxies by design, we interpret the trend as the  sky footprint selection effect of the \rdmp algorithm. To our relief, the stacked light profile of random points is consistent with 0 after the subtraction procedures. Using the jackknife method, we estimate the $1\sigma$ surface brightness uncertainties to be $\sim$31.5 mag arcsec$^{-2}$ at $r=1~\mathrm{arcmin}$ and $\sim$33 mag arcsec$^{-2}$  at $r=7~\mathrm{arcmin}$, which means that our methods are sensitive enough to detect light above these surface brightness levels.

Notably, the noise fluctuations of the random points appear to be highly correlated at different radii, as shown in Figure~\ref{fig:random_profile}. We notice that any image imperfections, such as scattered light, uneven sky-subtraction residual in different exposures, bright stars, etc., are likely to affect a significant portion of a single-exposure image (see the Appendix for some examples), causing correlated noise in the stacked profiles. The correlation is especially prominent at small radii, due to the smaller area. However, because the correlated fluctuations appear to be within the 3$\sigma$ uncertainty range in the random test (Figure~\ref{fig:random_process} bottom panel), we do not consider it further in this paper.

To test the accuracy of the ICL measurement using our method, we inject a simulated ICL+CG distribution into the images of the random points. The ICL+CG radial distribution is simulated as a combination of three Sersic profiles with parameters described in Section~\ref{sec:pure_icl}, and injected into the processed and masked images of the random points, which then go through the light profile estimation procedures described in Section~\ref{method:profile}. Poisson photon noise from the ICL+CG model is not included in the injection to create an ideal situation of low-noise ICL+CG observation. 

\begin{figure}
\includegraphics[width=0.5\textwidth]{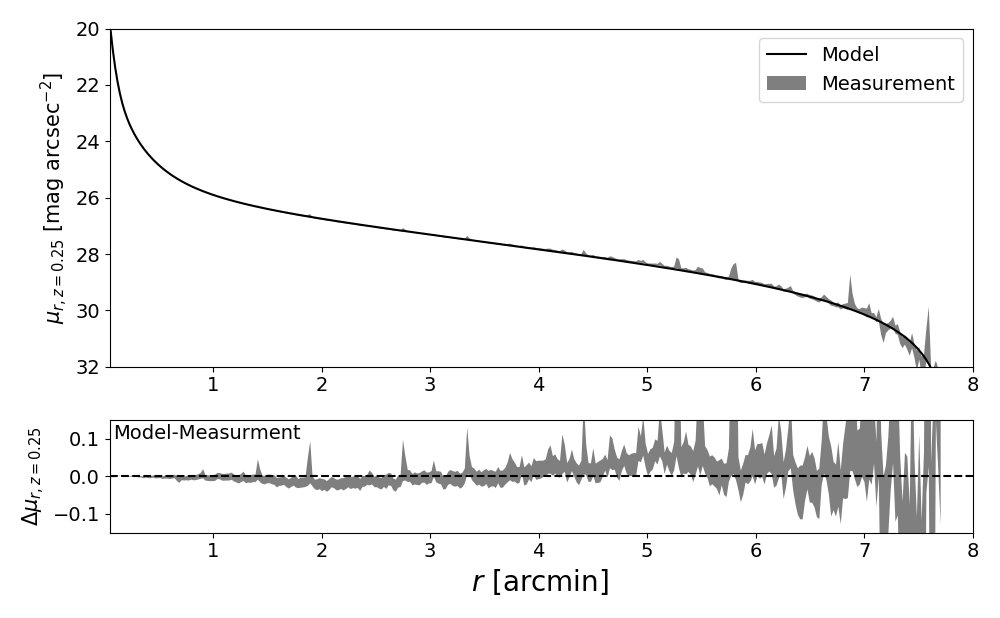}
\caption{Measurement of the simulated ICL+CG profile injected at the random points. The upper panel shows the measured ICL+CG profile, and the solid line shows the injected model. The method recovers the light profile of the simulated ICL distribution out to $\sim$ 7 arcmin from the center, at a surface brightness level of 30 $\mathrm{mag}/\mathrm{arcsec}^{2}$. The lower panel shows the residue between the measurement and the model, with the gray shaded area representing the uncertainties of the measurements.}
\label{fig:random_sim_test}
\end{figure}

The result of this test is presented in Figure~\ref{fig:random_sim_test}, showing the recovered radial profile of the simulated ICL+CG in comparison and the injected model. Because we assume that  the ICL flux reduces to 0 at $\sim$ 0\degree.145, the ICL+CG model shown in the figure has also been subtracted by its model values at $\sim$ 0\degree.145. Our method has recovered the injected ICL+CG model with excellent accuracy out to $\sim$ 7 arcmin from the injection center, down to a surface brightness level of 30  mag/arcsec$^{2}$.

\section{DECam Point-Spread Function (PSF)}
\label{sec:psf}
Literature studies show that the PSF of imaging instruments mounted on ground-based telescopes have low surface brightness wings that extend tens of arcminutes \citep{1969A&A.....3..455M, 1971PASP...83..199K, 1996PASP..108..699R, 2007ApJ...666..663B}. Detection of ICL is not to be confused with this feature. In this section, we measure and model the DECam \citep{2015AJ....150..150F} PSF in the $g-$ and $r-$ bands as a tension test of our methods, and also to determine whether or not scattered light from an extended PSF could have any significant effect on our detection of ICL.

To measure the central component of the PSF, we select a set of stars with magnitudes $19 < r < 19.3$, as measured by the DES \texttt{MAG\_PSF} quantity. These point-source objects are separated from galaxies using ``the modest classifier'' from the DES Y1 coadd catalog \citep{2017arXiv170801531D}. To further improve the purity of the point-source selection, we apply a color cut $\texttt{MAG\_PSF\_R} - \texttt{MAG\_PSF\_I} > 1.5$ as recommended in \cite{2015ApJ...807...50B}. This set of stellar objects is below the saturation limit of the DES single exposures in $g$ and $r$, and is thus useful for quantifying the central component of the PSF inside 10\arcsec.  We derive the stacked light profiles of these stellar objects as well as their uncertainties using methods outlined in Section~\ref{sec:methods}.

To study the farthest reaches of the DECam PSF,  we use a set of very bright stellar objects selected from the {\it Gaia} \footnote{https://www.gaia-eso.eu} database \citep{2016A&A...595A...2G}. The first set contains 57 bright stellar objects of $\sim 8$ magn in the {\it Gaia} $G-$band. Although these objects are saturated in DES images in the central 3\arcsec, they give high S/N when studying the outskirts of the PSF beyond 20\arcsec.  Finally, to derive the PSF radial profile between 10\arcsec and 20\arcsec, we select a set of moderately bright stellar objects which contains 100 stars of $\sim$14 mag in the {\it Gaia} $G-$band.  These objects are also saturated in their cores in DES images, but still have sufficient S/N at the radii where the unsaturated stars have no signal left above the noise level.  To compute the light profiles of these bright objects, we adopt the same image-processing procedure as described in Section~\ref{sec:methods}, and then a similar profile estimation procedure excluding the subtraction of random point profiles and the requirement of no bright stars within 526$\arcsec$. The light profile uncertainties of the bright star are estimated with a bootstrap approach by randomly sampling the stars.

\begin{figure}
\includegraphics[width=0.5\textwidth]{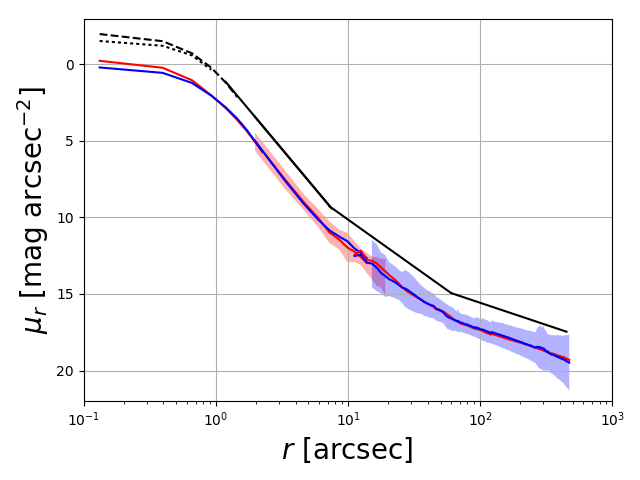}
\caption{Extended PSF of DECam+Blanco 4m in the $g-$ and $r-$ bands. The PSF is constructed by piecing together images of 19th, 14th, and 8th magnitude stars, each of which samples the PSF at good S/N (without saturating) at a different range of radii.  $g-$ and $r-$ band average profiles are shown with solid blue and red curves, respectively. The shaded red curve at $10\arcsec < r < 20\arcsec$ is derived with 14th magnitude $r-$band stellar images, and the shaded blue curve beyond 20\arcsec is from 8th magnitude stellar profiles.  This plot shows the profile with the central surface brightness normalized to  0 mag arcsec$^{-2}$. The black curves offset above the measured profiles are fitted models containing a Gaussian (dotted line, $g-$band)/Moffat(dashed line, $r-band$) central component and then three broken power-law sections as described in the Section~\ref{sec:psf}.}
\label{fig:decampsf}
\end{figure}

Figure \ref{fig:decampsf} shows the radial profiles of the PSF in the $g-$ and $r-$ bands for a composite of faint, intermediate, and bright stars.  The PSF is well characterized as a Gaussian of FWHM = 1.15\arcsec (FWHM = 1.043 \arcsec for $r$, and 1.18 \arcsec for $g$) at $r < 1.1$\arcsec, and by a three-component broken power law, $f \approx r^{-4.13}, ~\> 1.1\arcsec < r < 7.5\arcsec, ~f \approx r^{-2.46}, ~7.5\arcsec < r < 60\arcsec$, and $f\approx r^{-1.16}, ~60\arcsec < r < 450\arcsec$.   The effective seeing was slightly different for the $g-$ and $r-$ bands, yielding a slight difference in profiles in the inner arcsecond. A Moffat profile fits the inner profile to somewhat larger radii (1.5\arcsec) versus a Gaussian (1.1\arcsec). The red shaded band gives a typical profile range for a 14th magnitude $r$-band star at intermediate radii, and the blue shaded band gives the typical spread in $g$-band scattered light at larger radii. The slopes are identical in the $g$ and $r$-bands within the uncertainties.

With these derived PSF profiles, for a point source, we estimate that about 80\% of the light is contained within a radius of 1\arcsec, 98.3\% within 5\arcsec, 99.6\% within 50\arcsec, and 99.9\% within 400\arcsec.  Note that for the typical clusters measured here, 1 arcsec = 3.943 kpc (at  redshift 0.25). 
In section~\ref{sec:psf_discussion}, we further discuss the PSF effect on the interpretation of ICL results. 

\section{ICL+CG Profile} 
\label{sec:ICL}

\begin{figure}
\includegraphics[width=0.5\textwidth]{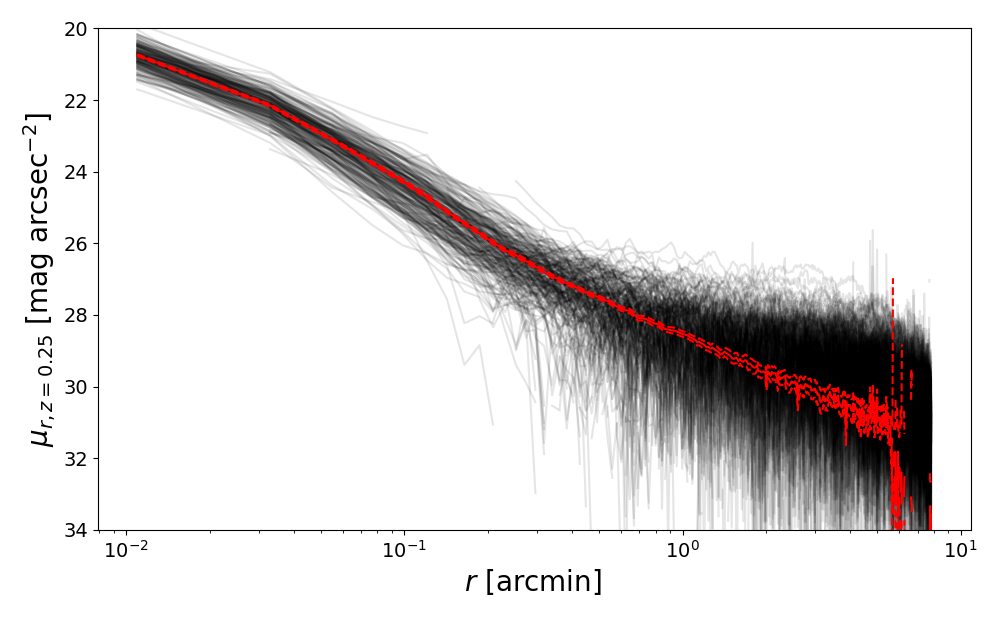}
\caption{This figure shows the ICL+CG profile measurements of individual clusters (black lines) and the stacked result (red solid and dashed lines indicating the mean and the uncertainties). Stacking significantly improves the precision of cluster light profile measurement beyond 0.5 arcmin.}
\label{fig:cluster_profile}
\end{figure}

In figure~\ref{fig:cluster_profile}, we show the derived light profiles of the individual clusters as well as their stacked light profile. These are the raw ICL+CG profile measurements in this analysis. Although on small scales within 0.5$\arcmin$ of the cluster centers the individual clusters show significant signals of ICL+CG, stacking enables the light profile measurement of ICL at distances beyond $\sim 0.5 \arcmin$ out to $\sim 6\arcmin$. In the rest of this section, we consider the residual light contribution from cluster galaxies and derive results such as ICL+CG color and ICL+CG integrated luminosity. Although this paper is aimed at analyzing the properties of ICL, we do not attempt to quantitatively dissect ICL and CGs in the analyses. We qualitatively refer to the diffuse light outside a few kpcs of the CG center as the ICL, and apply a ICL/CG radius cut when making quantitative evaluations.

\subsection{``Pure'' ICL+CG profile}
\label{sec:pure_icl}

With the procedures described in Section~\ref{sec:methods}, we derive the ICL+CG light profiles of each cluster. Unfortunately, not all of the light in the raw profile is associated with the ICL or CG. The light from the cluster foreground and background structures has been accounted for with the random point subtraction process in Section~\ref{sec:methods}. Faint, unmasked cluster galaxies and the cluster galaxy light leaking outside the masking apertures constitute a cluster light residual in the raw ICL+CG measurements. We estimate the residual from the light profile of cluster galaxies.

We first derive the stacked light profile of non-CG cluster galaxies (cluster satellite galaxies). The methods described in Section~\ref{sec:methods}, including the background residual and random light profile subtractions, are applied to the same set of clusters, but masking only objects brighter than 18 mag. Subtracting the previously derived light profiles with masking to 22.4 mag gives the radial light profile of non-CG cluster galaxies between 18 and 22.4 mag.

For each of the clusters, we assume the luminosity of non-CG cluster galaxies (cluster satellite galaxies) to follow a single Schechter function \citep{1976ApJ...203..297S}, and derive (1) the luminosity fraction between the 18 and 22.4 mag range and (2) the luminosity fraction beyond 22.4 mag. For the Schechter function parameters, we assume a faint-end slope of -1.0 and the characteristic magnitude of 1.29 $L_\odot/h^2$ at redshift 0.25 \citep{2009ApJ...699.1333H}. 
We further assume that the light profile of cluster satellite galaxies in different luminosity ranges have the same radial distributions (same shape of the radial light profiles). With the above luminosity fraction estimates and the measurement of non-CG cluster galaxy light profile between 18 and 22.4 mag, we derive the light profiles of cluster satellite galaxies and the light profile of satellite galaxies fainter than 22.4 mag. The cluster satellite galaxies below 22.4 magnitude makes up $\sim$ 4\% of the total cluster satellite galaxy light in the redshift range of 0.2-0.3.

Our masking apertures (2.5 Kron radii) may not fully enclose the light of cluster galaxies due to their extended radial profiles. \cite{2005PASA...22..118G} estimated that the unmasked component makes up  $0.1 - 9.6\%$ of the S\'ersic \citep{1963BAAA....6...41S} galaxy profiles. We aggressively assume the masking residual to be 9.6\% of the cluster galaxy light. From the raw ICL profile of each cluster, we subtract the light of faint, unmasked cluster satellite galaxies below 22.4 magnitude and the residual light of the masked galaxies assuming that they both have the same radial distribution as the general cluster satellite galaxy population. 

We correct the derived ``pure'' ICL+CG light profiles for wavelength shifting (K-correction) and passive redshift evolution (e-correction)  to an observer redshift of 0.25 using a stellar population template\footnote{http://www.baryons.org/ezgal/} \citep{2003MNRAS.344.1000B, 2012PASP..124..606M} with a single starburst of metallicity $Z=0.008$ at $z=3.0$. These corrected profiles are interpolated onto a physical radius grid and then averaged to compute a stacked ``pure'' ICL+CG profile, using the jackknife method.

\begin{figure*}
\includegraphics[width=1\textwidth]{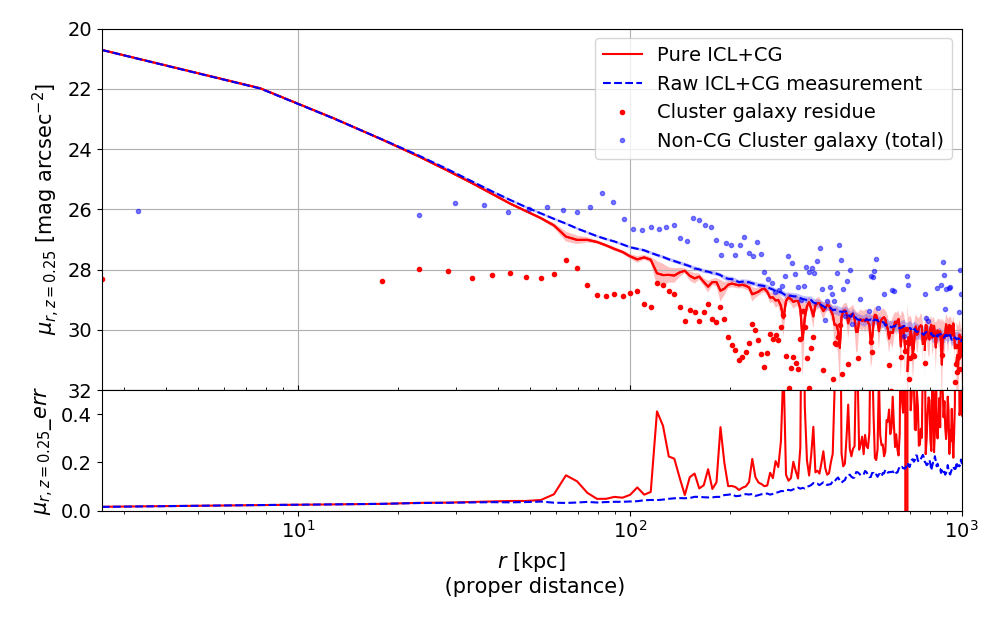}
\caption{This figure shows the derived ICL+CG profiles (upper panel) and the uncertainties of the measurements (lower panel). The cluster galaxy residual (red point) is subtracted from the raw ICL+CG measurement (blue line) to derive the "pure" ICL+CG profile (red line). Uncertainties of the profiles are displayed as the shaded regions (upper panel) and also shown in the lower panel. The ICL+CG profiles are measured with high S/N to 1 Mpc, although the subtraction of cluster galaxy residual introduces significant noise (poisson noise) into the "pure" ICL profile.}
\label{fig:cluster_profile2}
\end{figure*}

\begin{figure}
\includegraphics[width=0.5\textwidth]{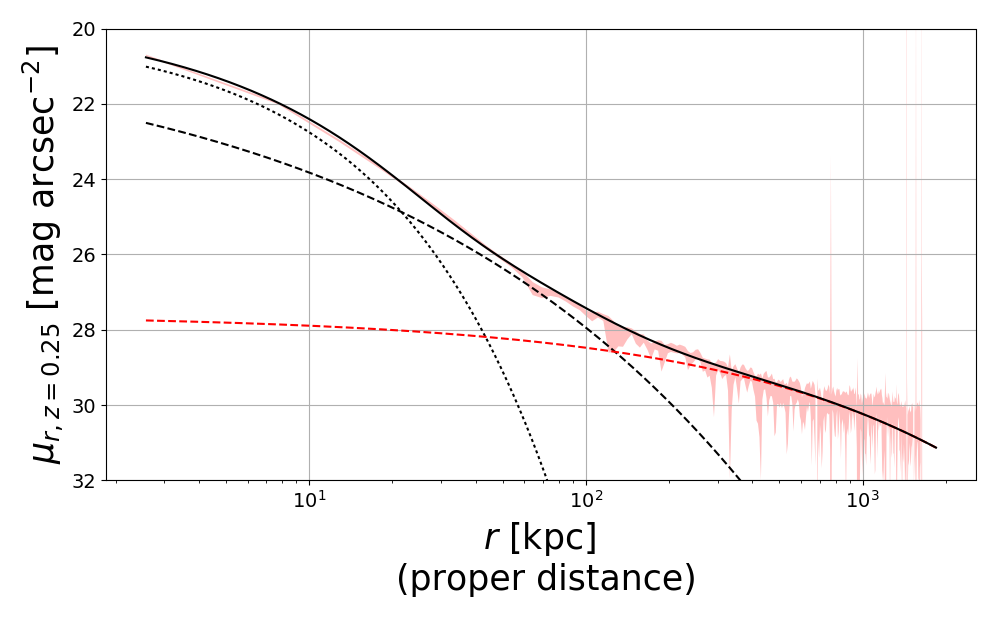}
\caption{The ICL+CG light profile can be approximated with three Sersic components (black solid line): a core disk component that is dominant within 10 kpc (dotted line), a bulge component that is dominant between 30 and 100 kpc (dashed line), and a diffuse component that is dominant outside 200 kpc (red dashed line).}
\label{fig:profile_fit}
\end{figure}

The radial profiles of ICL+CG are shown in Figure~\ref{fig:cluster_profile2}. The raw  measurement is dominated by pure ICL+CG light to $\sim 100$ kpc, beyond which the cluster galaxy residual becomes relevant. The subtraction of the cluster galaxy residual introduces significant noise to the pure ICL+CG detection. Nevertheless, the stacked ICL+CG profile is measured with high S/N to 1 Mpc, reaching 30 mag arcsec$^{-2}$.

\begin{table}
\caption{Fitted parameters of the ICL $r-$band radial profile to Sersic models, $I(r)=Ie\times\mathrm{exp}(-b_n(r/Re)^{1/n})$.}
\label{tbl:sersic_fits}
\vspace{1em}
\begin{tabular}{lcccc}
\hline
 & $Ie^{*}$  & $n$  & $Re$\\
  & (flux/arcsec$^2$)  &   & \\
\hline
First Sersic Component & $9830\pm162$ & $1.34\pm0.023$ & $9.13\pm0.24$ kpc\\
  \hspace{0.5em} \\
\\
Second Sersic Component & $8846\pm1046$ & $3.07\pm0.08$ & $52.1\pm2.2$ kpc\\
  \hspace{0.5em} \\
\\
Third Sersic Component & $9.1\pm3.3$ & $2.1\pm0.4$ & $2.6\pm0.7$ Mpc\\
  \hspace{0.5em} \\
  \hline
\\
\end{tabular}
$^{*}$ $Ie$ has been scaled so that the magnitude zero point of $I(r)$ is 30.
\end{table}

Similar to the finding in \cite{2018AstL...44....8K}, the ``pure'' ICL+CG radial profile is best described with a combination of three Sersic models (Figure~\ref{fig:profile_fit}): a dominant core of Sersic index $1.34\pm0.023$ and radius $9.13\pm 0.24$ kpc within the inner 10 kpc, a bulge between 30 and 100 kpc with Sersic index $3.1\pm0.08$ and radius $52.1\pm 2.2$ kpc, and a diffuse component dominant outside 200 kpc with index $2.1\pm0.4$ and radius $2.6\pm 0.7$ Mpc. The fitted Sersic profile parameters are listed in Table~\ref{tbl:sersic_fits}.

\subsection{ICL+CG Integrated Luminosity}
\label{sec:integration}

\begin{figure}
\includegraphics[width=0.5\textwidth]{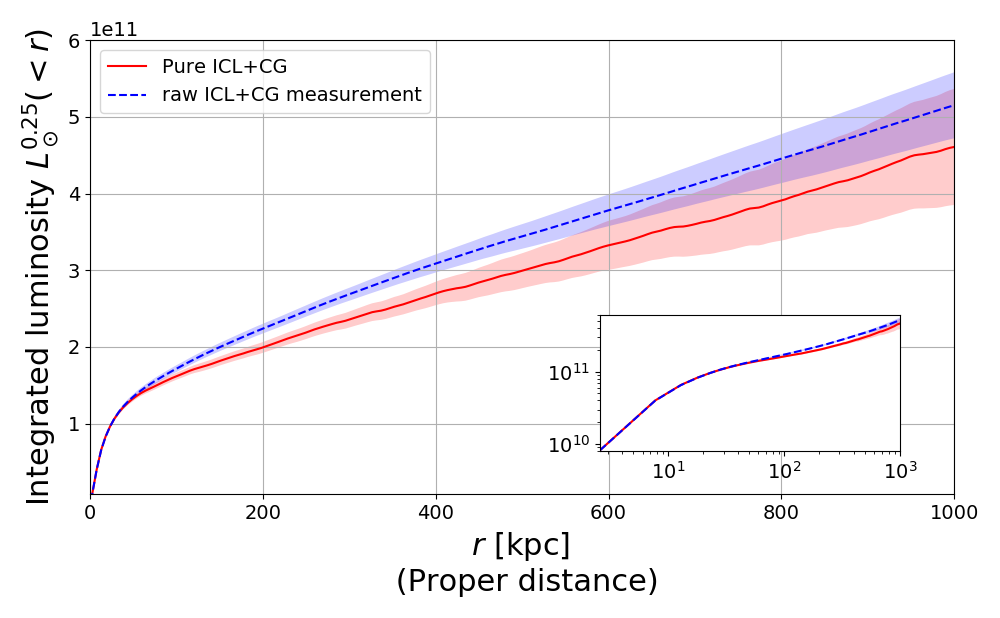}
\caption{Integrated ICL+CG luminosity within different radii. The inset figure shows the same plot but with log scaling. Within $\sim30$ kpc, the integrated ICL+CG luminosity steeply increases with distance because of the dominance of the CGs. The increase slows down outside $\sim$ 30 kpc, indicating that the light distribution has transitioned into a diffuse component.}
\label{fig:cluster_profile3}
\end{figure}

\begin{figure*}
\includegraphics[width=1.0\textwidth]{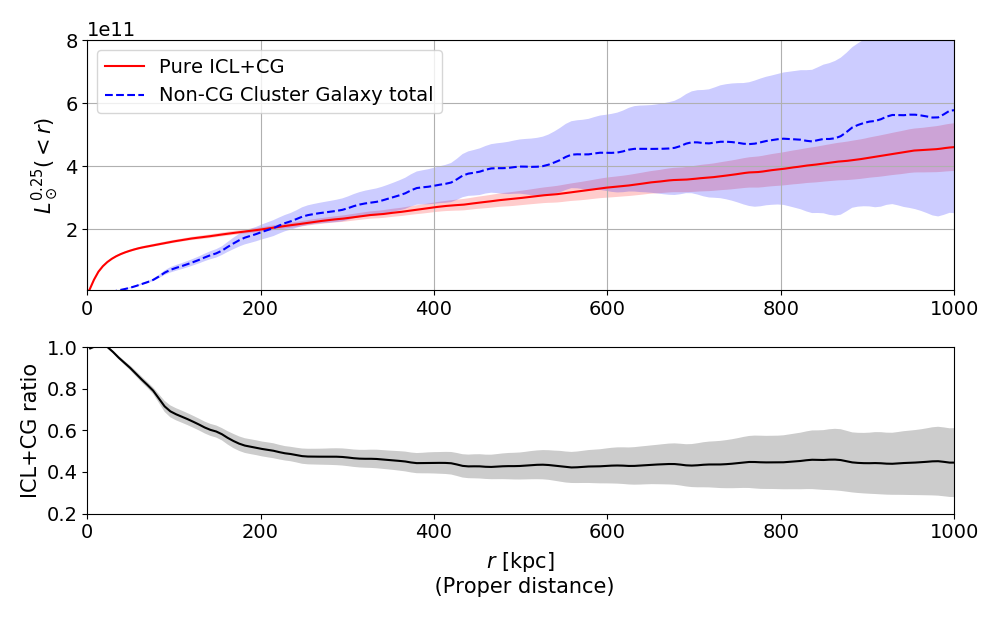}
\caption{Integrated ICL+CG luminosity (red) and non-CG cluster galaxy luminosity (blue) within different radii. Within the inner 200 kpc, ICL+CG luminosity is more abundant than that of the non-CG cluster galaxies. Outside 200 kpc, the total luminosity of  non-CG cluster galaxies becomes higher. Within 1 Mpc, the luminosity of ICL+CG makes up 44\% (bottom panel)  of the total cluster stellar luminosity computed as the sum of the non-CG cluster galaxies and the ICL+CG. }
\label{fig:cluster_profile4}
\end{figure*}

A significant fraction of the cluster stellar luminosity may be contained in ICL+CG, which is a debated topic in the literature. With the ICL+CG surface brightness profile derived in the previous sections, we examine the integrated luminosity profile of ICL+CG at different radii as
\begin{equation}
f(r)= 2\pi \int_0^{r} r' l(r') \mathrm{d}r'.
\end{equation}
 $l(r')$ is the ICL+CG surface luminosity profile from the previous subsection, corrected to an observer frame of redshift 0.25. We compute the integrated ICL+CG luminosity for both the ``raw" and ``pure" ICL+CG radial profiles, the results of which are shown in Figure~\ref{fig:cluster_profile3}.

These integrated ICL+CG luminosity profiles are measured with high S/N out to 1 Mpc. Within $\sim$ 30 kpc, the integrated luminosity increases steeply with radius because of the contribution from the cores of the CGs. Outside $\sim 30$ kpc, the increase slows down, indicating that the light profile has transitioned into a diffuse component. 

The integrated luminosity within CGs and ICL is an important quantity in understanding CG and ICL formation. For example, \cite{2016ApJ...816...98Z} speculated that 30 to 60\% of the stars that merged into the CGs need to be deposited outside 32 kpc at redshift 0 in order to explain the redshift evolution of the CGs. We find that the ICL integrated luminosity between 32 and 200 kpc makes up $\sim$ 42\% of the total ICL+CG luminosity. The ICL integrated luminosity between 32 kpc and 1 Mpc is even more significant, $\sim$ 3 times as luminous as the luminosity enclosed within 32 kpc. 
Interestingly, the integrated luminosity of the ICL does not seem to converge within the radius range of 1 Mpc. For a more thorough consensus of ICL total luminosity, future studies may have to investigate an even larger radius range.

We also compute the integrated luminosity of the non-CG cluster galaxies (cluster satellite galaxies), using the light profile measurements of these galaxies in section~\ref{sec:pure_icl}. The derivation of this satellite galaxy luminosity is susceptible to modeling uncertainties of the  galaxy luminosity function and the galaxy-masking apertures. Nevertheless, the results are shown in Figure~\ref{fig:cluster_profile4}, in comparison with the integrated luminosities of ICL+CG.

Within 200 kpc, the integrated luminosity of ICL+CG surpasses the total luminosity of cluster satellite galaxies (non-CG cluster galaxies). Outside 200 kpc, the satellite galaxy luminosity becomes paramount to that of ICL+CG. We further estimate the ratio of ICL+CG in the total cluster stellar luminosity, which is computed as the sum of ICL+CG and non-CG cluster galaxies. We refer to this quantity as the ICL+CG fraction, as in \cite{2015MNRAS.449.2353B}. The ICL+CG fraction gradually drops when we enlarge the radius of the luminosity integration, reaching $44\pm17$\% at 1 Mpc.

\subsection{ICL+CG color}

\label{sec:color}

\begin{figure}
\includegraphics[width=0.5\textwidth]{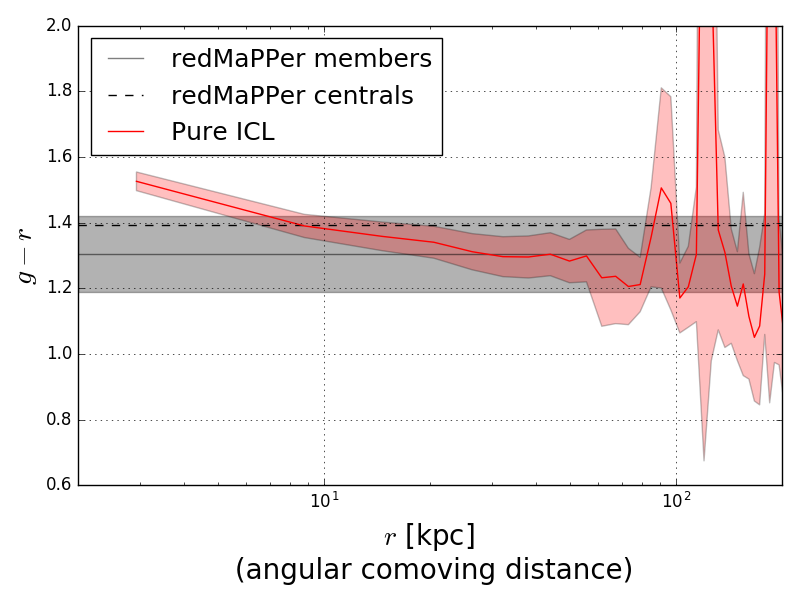}
\caption{$g-$ and $r-$band radial profiles (upper panel) and color profile of pure ICL+CG (lower panel). For comparison, we also show the average colors of the CGs and satellite red sequence members (K-corrected to $z=0.25$), selected with the redMaPPer algorithm, and the ICL+CG colors derived in Zibetti et al. 2005. The shaded regions represent the uncertainties of the color measurements.}
\label{fig:color}
\end{figure}

Using our measurements of the ICL+CG profiles in both the $g-$ and $r-$bands, we further derive the ICL+CG color profile (Figure~\ref{fig:color}). The $g-$band ICL+CG measurement has a greater level of noise compared to the $r-$band measurement (Figure~\ref{fig:color} upper panel), because of less accurate estimation of cluster galaxy residual. Because the color measurements require high signal to noise flux measurements in both the $g$ and $r$ bands, our color derivation is only robust to 90 kpc from the center. Outside 90 kpc, the color measurements are dominated by noise. 

We compare the ICL+CG colors to those of the cluster CGs and the satellite red sequence galaxies selected by redMaPPer. These galaxy colors are derived with the \texttt{MODEL\_MAG} magnitudes in the DES database \citep{2017arXiv170801531D}, K-corrected to $z=0.25$ as is done for the ICL+CG profiles. Overall, the core  component of the ICL+CG profile appears to be consistent with the CG measurement in the DES database,  but the ICL color beyond $\sim 20$ kpc is more consistent with the average color of the satellite red sequence galaxies. The ICL+CG color also displays a radial trend of becoming bluer at a larger radius. We find that the color trend in the central 10 kpc of the ICL+CG combination is likely caused by the PSF differences in the $g-$ and $r-$bands (see section~\ref{sec:psf_discussion} for detailed discussions). Between 10 and 90 kpc, a $\chi^2$ minimization gives a significant radial gradient,
\begin{equation}
\nabla (g-r) \equiv \frac{{\rm d} (g-r)}{{\rm d \, Log}(r)}=-0.152\pm 0.027,
\end{equation}
 indicating a robust color radial trend.

\section{Cluster Mass Dependence}
\label{sec:clusters}

\subsection{More ICL in Richer Clusters}

\begin{figure}
\includegraphics[width=0.5\textwidth]{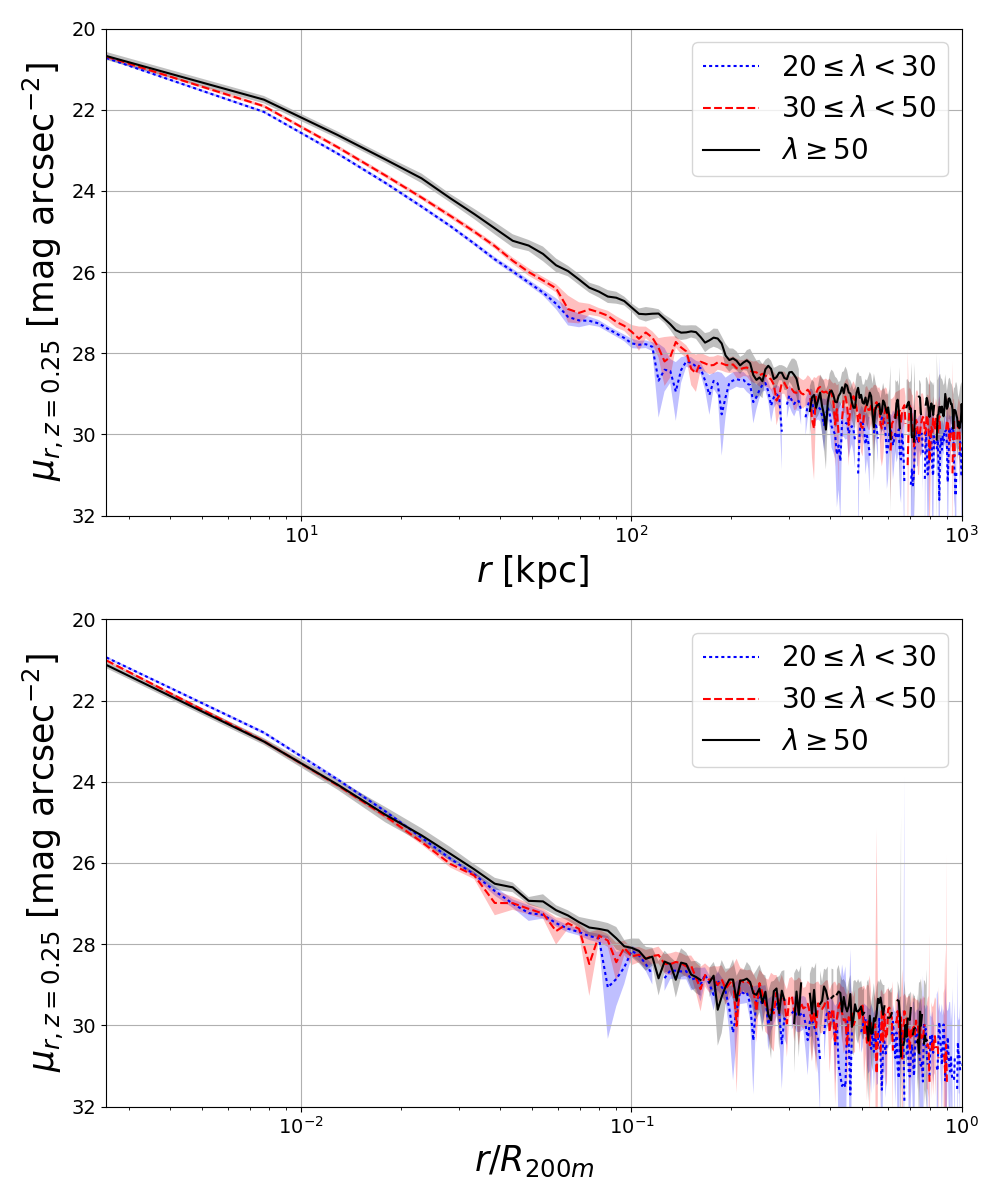}
\caption{Upper panel: ICL profiles of clusters in different richness ranges. The ICL is more luminous and extended in richer systems. Lower panel: ICL profiles appear to be ``self-similar'', in that after scaling with cluster radius ($R_{200m}$), the ICL profiles of clusters in different richness ranges are indistinguishable.}
\label{fig:lambda}
\end{figure}

\begin{figure}
\includegraphics[width=0.5\textwidth]{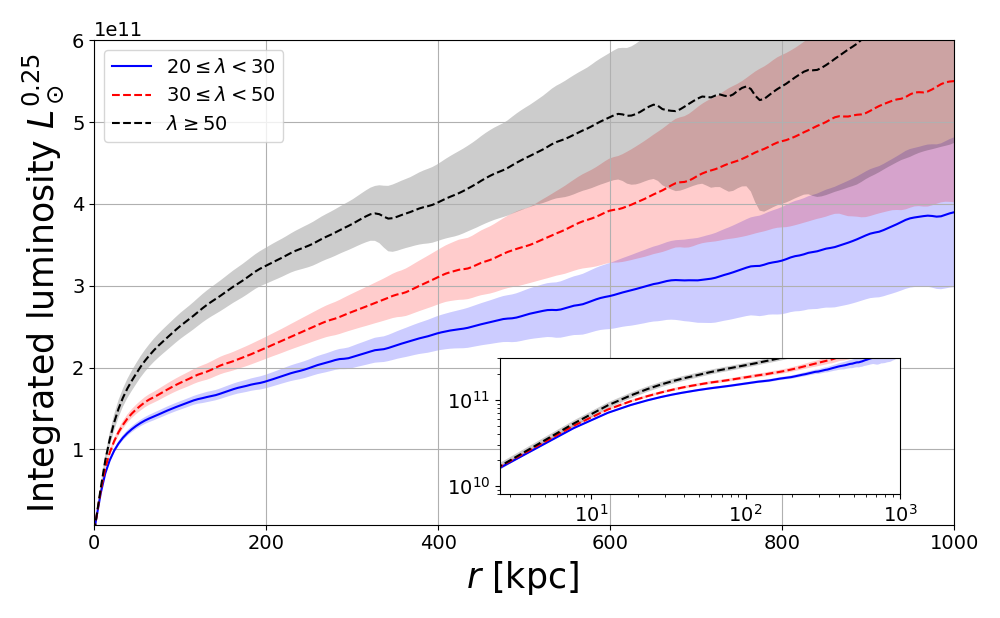}
\caption{Integrated ICL luminosities at different radii of clusters in different richness ranges. The inset figure shows the same plot but with log scaling. The integrated ICL luminosities appear similar within $\sim30$ kpc regardless of cluster richness. As the distance increases, richer clusters contain more ICL, which means that the ICL dependence on cluster richness is mainly driven by the diffuse light outside the cores of the CGs.}
\label{fig:cluster_profile3_lambda}
\end{figure}

The \rdmp algorithm adopts richness, denoted by $\lambda$, as a mass proxy, which is defined as the number of cluster red sequence galaxies above a luminosity threshold. Using this richness quantity as a cluster mass proxy, we examine the variation of ICL surface brightness with mass.

The $\sim$ 300 clusters studied in the paper are further divided into three subsamples, $20\le \lambda < 30$, $30\le \lambda < 50$, and $\lambda \ge 50$, representing three cluster mass ranges with mean masses of $10^{14.09}$,  $10^{14.30}$, and  $10^{14.61}$ $\mathrm{M_\odot}/h$ \citep{2017MNRAS.469.4899M}. We derive the stacked ICL+CG profiles of these three subsamples as shown in Figure \ref{fig:lambda}. The ICL is more luminous in richer and hence more massive clusters. The core components within $\sim$ 10 kpc of the ICL+CG profile appear similar among the three cluster richness subsamples, in agreement with the inside-out growth scenario \citep[e.g., ][]{2010ApJ...709.1018V,2015A&A...577A..19V} wherein the cores of the CGs form early and the accreted CG stellar content is deposited onto the outskirt at a later time. The "inside-out" trend is further demonstrated in the integrated luminosity profiles (Figure~\ref{fig:cluster_profile3_lambda}). Richer clusters are more abundant in the total ICL+CG luminosity, but the differences start outside 20 kpc and rapidly enlarge at larger radii.

\begin{figure}
\includegraphics[width=0.5\textwidth]{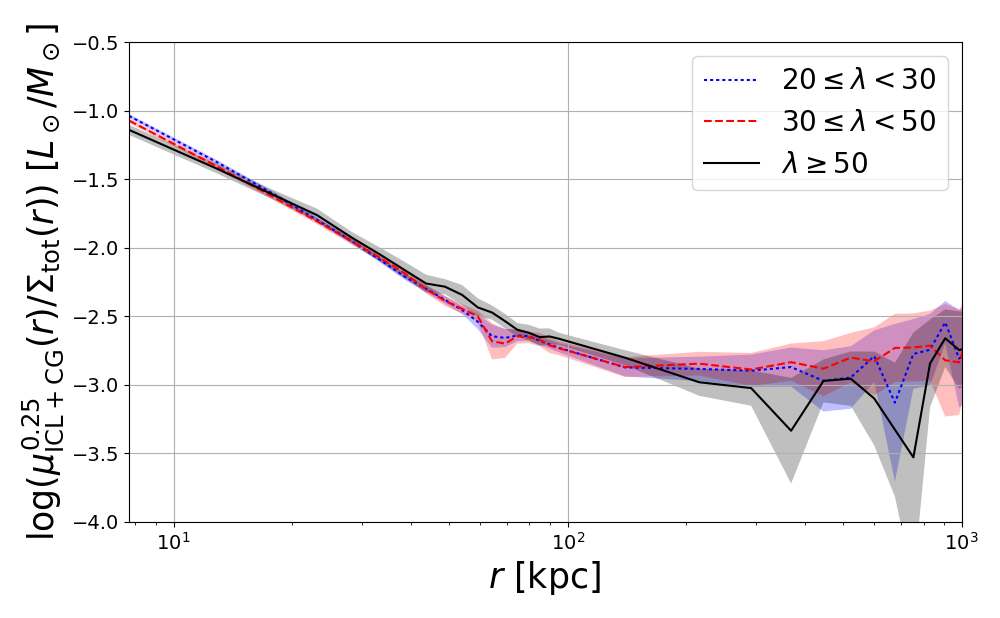}
\caption{Ratios between ICL+CG luminosity and total cluster mass at different radii. Outside 100 kpc, the ICL luminosity-to-cluster-mass ratio appears flat within uncertainties, an indication that ICL luminosity traces the cluster mass distribution. Interestingly, the radial range of ratio flattening is coincident with where the ICL+CG profile transitions into a diffuse component (Figure~\ref{fig:profile_fit}). Note that the ICL+CG profiles have been rebinned by a factor of 15 outside 100 kpc from Figure \ref{fig:lambda} to reduce noise.}
\label{fig:lambda_sigma}
\end{figure}

\subsection{ICL+CG and Cluster Mass Distribution}

Interestingly, after scaling the ICL+CG profiles by the radii of clusters, the radial profiles of ICL+CG in different richness ranges appear similar (Figure~\ref{fig:lambda}). In this exercise, instead of stacking together ICL+CG radial profiles in terms of physical distances, we stack them in radial bins of $r/R_{200m}$, with $R_{200m}$ derived from  the mass-richness relation in \cite{2017MNRAS.469.4899M}. This is the first evidence of ICL distribution being ``self-similar", as in the self-similarity of cluster mass or gas distributions \citep[e.g., see a review in][]{2012ARA&A..50..353K} wherein their radial profiles are indistinguishable after scaling with a characteristic radius.

We further examine the resemblance between the ICL and cluster mass distributions, by comparing their radial profiles. The state-of-the-art constraint of cluster mass distribution (including dark and baryonic matter) comes from weak lensing studies, yet cluster surface mass profile is not a direct observable from weak lensing\footnote{Weak lensing directly measures excess matter density enclosed within a radius, rather than the surface mass density at that radius.}, and its reconstruction can be noisy. In this comparison, we use a cluster surface mass density model from \cite{2018arXiv180500039M} \footnote{\url{https://tmcclintock.github.io/code/cluster\_toolkit}}, which is a combination of a Navarro-Frenk-White (NFW)  model \citep{1997ApJ...490..493N} and a two-halo matter correlation model, projected onto the plane of the sky: 
\begin{equation}
\begin{split}
\xi(r|M, c) &=\mathrm{max}(\xi_{NFW}(r|M, c), \xi_{2h}(r|M)), \\
\Sigma(r|M, c, z) & = \int^{\infty} _{-\infty} d \chi \rho_m(z) \xi(\sqrt{r^2+\chi^2}|M, c).
\end{split}
\end{equation}
In the above equations, $\xi_{NFW}(r|M, c)$ and $\xi_{2h}(r|M)$ represent the cluster matter correlation functions in the one-halo and two-halo regimes of a cluster with mass $M$ and concentration $c$. $\rho_m(z)$ represents the universe mean matter density at redshift $z$. $\Sigma(r|M, c, z)$ is the surface mass density at distance $r$ on the plane of the sky, integrated along the line-of-sight distance $\chi$. We compute $\Sigma(r|M, c, z)$ at the mean masses of the three cluster richness subsamples, $10^{14.09}$,  $10^{14.30}$ and  $10^{14.61}$ $\mathrm{M_\odot}/h$, assuming the concentration to be 5 and the redshift to be 0.25.

The ratio between the ICL+CG luminosity and the cluster surface mass density is shown in Figure~\ref{fig:lambda_sigma}. This luminosity-to-mass ratio is at its highest within 100 kpc. As found of the cluster matter distribution in multiwavelength observations \citep{2013ApJ...765...24N, 2013ApJ...765...25N}, the core of galaxy clusters is dominated by the stellar mass of the CGs, which would explain the high ICL to cluster mass ratio near the center. 

The ratio drops with enlarging radius. Between 100 kpc and 1 Mpc, the ICL luminosity-to-cluster-mass ratio no longer displays a noticeable radial trend, which is $10^{-2.85\pm0.09}$ at $r=215$ kpc and $10^{-2.84\pm0.4}$ at $r=985$ kpc, in the medium richness bin. Because the cluster surface mass density \citep{2018arXiv180500039M} and the ICL surface brightness (Section~\ref{sec:pure_icl}) both drop by a factor of $\sim 10$ in this radial range, the flatness of the ICL+CG luminosity-to-mass ratio indicates that the ICL luminosity distribution closely follows the cluster mass distribution. Moreover, throughout the 100 kpc to 1 Mpc radial range, the ICL+CG luminosity to cluster mass ratios are similar among the three cluster richness subsamples, a further evidence that the ICL+CG luminosity profile is self-similar and traces the overall cluster mass distribution.

The 100 kpc to 1 Mpc radial range is also coincident with where the ICL+CG profile transitions into an extended diffuse component (Figure~\ref{fig:profile_fit}). Thus, this transition potentially marks a physical separation of ICL and CG.

\section{Systematics}
\label{sec:systematics}
\subsection{Sky Subtraction}
\label{sec:sky_effect}

\begin{figure}
\includegraphics[width=0.45\textwidth]{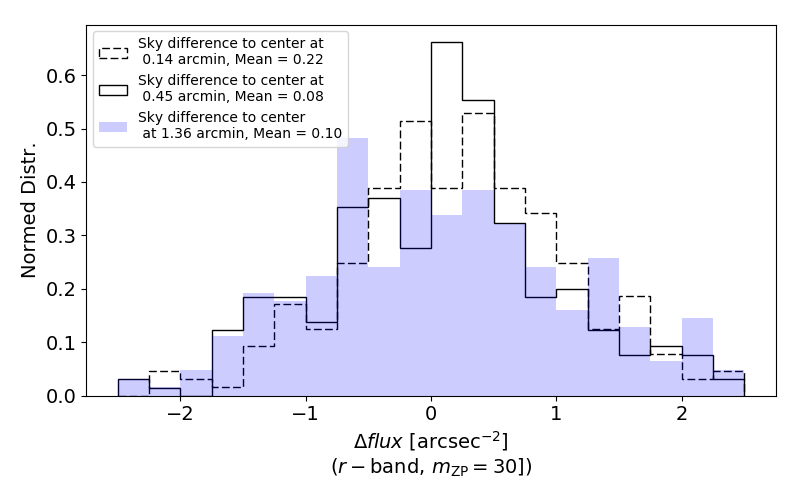}
\caption{The differences in the sky level estimations between the cluster centers and locations away from the cluster centers.  These difference are calculated by subtracting the sky level at a distance from the sky level at the cluster center, rescaled as the flux per arcsec$^2$ with a magnitude zero point of 30. See section~\ref{sec:sky_effect} for discussions on the implications of this figure.}
\label{fig:sky_effect}
\end{figure}

Sky subtraction near bright galaxies or stars is a known difficulty in processing imaging data. It is common for the processing software or pipelines to overestimate the sky level around bright objects because of light contamination from the objects, and hence oversubtract the sky background. This kind of effect is particularly detrimental to the detection of low surface brightness light near bright galaxies, and in our case, the detection of faint diffuse light around the CGs. 

In this paper, the sky levels are estimated and subtracted during three processes:
\begin{enumerate}
\item Sky level is estimated for each exposure image and subtracted before the single-exposure images are coadded.
\item In section~\ref{method:profile}, residual sky background is estimated and subtracted again by sampling the light profiles of random points that cover the footprint of the cluster sample to account for cluster selection effects (i.e. we only look for clusters in sky regions that are not in immediate proximity to bright stars or nearby galaxies).
\item In section~\ref{method:profile}, residual background light is again estimated as the flux level at a distance of $\sim$ 7 arcmin (about $\sim$1.6 Mpc at redshift 0.25) from the cluster center and  then subtracted to reduce the cluster-to-cluster variations of the ICL detection.
\end{enumerate}
In the above steps, (2) and (3) are by definition either far away or not correlated with cluster locations, hence we do not consider them to be contaminated by the light of the  CGs. As previously mentioned in Section~\ref{sec:finalcut}, the sky level estimation in step (1) is derived through modeling the variation of light over the 3 deg$^2$ DECam focal plane and thus should be relatively insensitive to the presence of individual bright galaxies or stars. 

Nevertheless, we investigate whether or not the sky level in step (1) may have been overestimated near the cluster center. In this exercise, we look into the sky background maps estimated for the single-exposure images covering the \rdmp CGs, and directly compare the sky values estimated  at the location of the cluster center and at locations away from the center. 

For each of the $\sim$ 300 clusters analyzed in the paper, we search for single-exposure images that cover at least the central $2.06 \times 2.06$ arcmin$^2$ region of the cluster\footnote{The size of each DES CCD image is approximately $8.77 \times 17.5$ arcmin$^2$, and varies slightly depending on the CCD location on the focal plane.}. Because of the survey nature of DES, the central region of each cluster may not fully appear within one exposure image, and we would like to avoid comparing sky levels in different exposures as they may vary significantly. For those single-exposure images that fully overlap with the cluster's central $2.06 \times 2.06$ arcmin$^2$ regions, we compute the sky flux difference between the cluster center location and locations at distances away from the centers, with the differences rescaled to a magnitude zero point of 30 for each exposure. 

Figure~\ref{fig:sky_effect} shows the sky value differences between the cluster center and  locations that are at distances of 0.14, 0.45, and 1.36 arcmin, corresponding to physical distance separations of 33, 106, and 320 kpc at $z=0.25$. If the sky level is indeed overestimated at the cluster center, we would expect the differences to be shifted to the positive side, and the shifts may become larger at larger distances. However, neither of these trends appear to be present in this test.  The mean of the differences is at a flux level of $\sim 0.1$ arcsec$^{-2}$ between the cluster centers and  locations 1.36 arcmin away from the centers, significantly under the ICL flux limit in this work (30 mag $\mathrm{arcsec}^{-2}$ at $\sim$ 1 Mpc corresponds to a flux of $\sim 1$ arcsec$^{-2}$ with a magnitude zero point of 30).

\subsection{PSF Effect}
\label{sec:psf_discussion}

\begin{figure}
\includegraphics[width=0.45\textwidth]{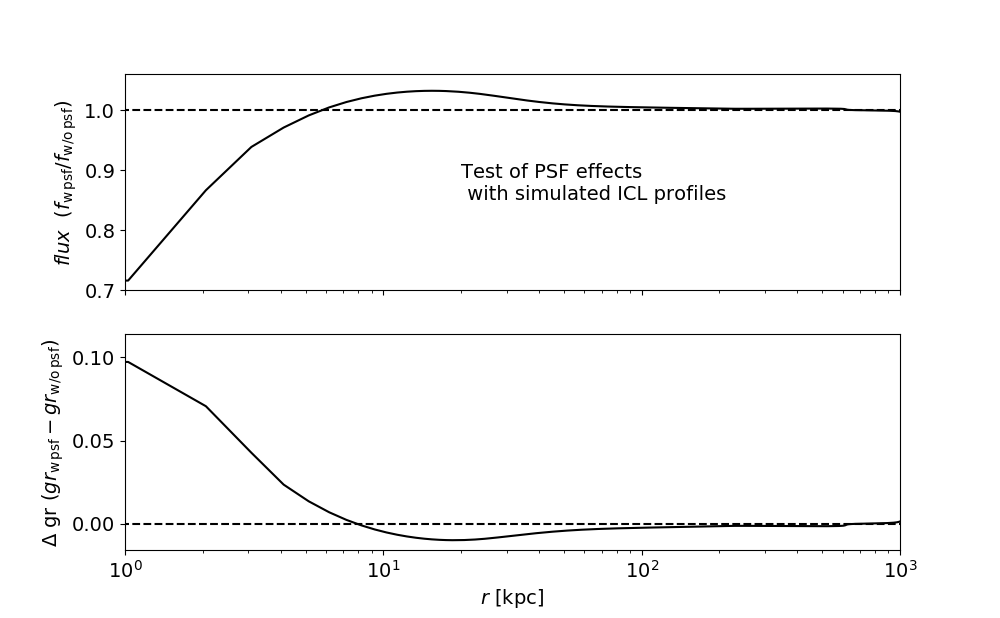}
\caption{The effects of the PSF on the measurement of the flux profile (upper panel) and color profile (lower panel) of ICL+CG, tested with a simulated ICL+CG profile. We create the simulated ICL+CG profile at $z=0.25$ using a combination of three Sersic models (Section~\ref{sec:pure_icl}), and convolve it respectively with the average DES Y3 $g$ and $r$ PSFs derived in Section ~\ref{sec:psf}. The upper panel shows the difference in flux per arcsec$^2$ before and after the PSF convolution in the $r$ band. The lower panel show the $g-r$ color difference before and after PSF convolutions in the $g-$ and $r-$bands. We find that the PSF shifts both the flux profile and the color profile, but only within the inner 10 kpc. See section ~\ref{sec:psf_discussion} for more detailed discussions.}
\label{fig:psf_effect}
\end{figure}

A frequently debated topic in the study of galaxy low surface brightness light envelope is the effect of PSFs. As noted in \cite{1969A&A.....3..455M, 1971PASP...83..199K, 1996PASP..108..699R, 2007ApJ...666..663B}, the telescope and instrument point spread functions almost always have an extended component that radially extends several arcseconds or arcminutes.  This is also shown to be true for the DECam images on the Blanco telescope in Section~\ref{sec:psf}. In previous studies about whether or not faint light envelope light exists around galaxies \citep{2014MNRAS.443.1433D, 2015MNRAS.446..120D} including the Milky Way and other spiral \citep{1994Natur.370..441S, 2004MNRAS.347..556Z, 2004MNRAS.352L...6Z, 2014ApJ...782L..24V, 2015ApJ...800..120Z} or elliptical galaxies \citep{2011ApJ...731...89T}, \cite{2008MNRAS.388.1521D, 2014A&A...567A..97S, 2015A&A...577A.106S} showed that the extended component of the telescope and camera PSF may partially or largely account for the observed galaxy light halo and its color gradient.

In this section, we consider whether or not the extended component of DECam/Blanco PSFs can explain the ICL measurements in this paper. The measurement of the extended wings of the PSF and ICL profiles do not have sufficient S/N to allow the smooth deconvolution of PSF from ICL. To evaluate the effect of PSFs, we creates a simulated ICL profile, convolve it with a PSF, and examine the difference in the ICL profiles before and after the convolution (Figure~\ref{fig:psf_effect}). The unconvolved ICL profile is modeled  with the three Sersic components as described in Section ~\ref{sec:pure_icl}, and we convolve it to the extended PSF model derived in Section~\ref{sec:psf}. We find that the PSF convolution only significantly alters the shape of the ICL within the inner 10 kpc, and is negligible outside 100 kpc at redshift 0.25.

To estimate the effect of PSFs on the ICL color measurements, we compute the color shifts when convolving the ICL profile model to the $g-$ and $r-$band PSF models from Section ~\ref{sec:psf}. Driven by the wider PSF FWHM in the $g-$ band compared to that of the $r-$band, PSF convolution produces an artificial trend of bluer color at a large radius within the inner 10 kpc, which informs us not to analyze the ICL color gradient in this radial region. The color shift caused by PSF convolution is negligible outside 10 kpc.

\subsection{Masking of Cluster Galaxies}

As discussed in Section~\ref{sec:pure_icl}, the raw ICL measurements contain light  from faint unmasked cluster galaxies, as well as residual light outside the masks of the already masked galaxies. These contributions are subtracted from the raw ICL measurements to derive pure ICL profiles, by assuming (1) a luminosity function of cluster galaxies and that (2) 9.6\% of the galaxies' light leaks to the outside of the masks.

As discussed in Section~\ref{sec:pure_icl}, we assume a single Schechter luminosity function to estimate the light residual from faint unmasked cluster galaxies. As with the parameters of the Schechter function, given recent measurement uncertainties of the cluster galaxy luminosity function \citep[e.g.,][]{2016MNRAS.459.3998L, 2018arXiv180703207R}, we find that varying the faint-end slope parameter by $\pm0.2$ or the characteristic magnitude parameter by $\pm 0.5$ mag  changes the estimated fraction of residual light by $\sim 1\%$. 

Notably, we have ignored possible deviations from a single Schechter function caused by faint dwarf cluster galaxies. \cite{2016MNRAS.459.3998L} estimated that the faint component of cluster galaxies has a normalization factor about 8\% of the brighter ones, with a characteristic magnitude $\sim 3.5$ mag fainter and a faint-end slope of $\sim -2.0$.  Such a faint dwarf galaxy component makes up less than 1\% of the total light of the bright component, and hence ignored in our analysis.

The second contamination to the ICL measurement, the residual light outside the masks of the already masked galaxies, is assumed to be 9.6\% of the total light of those galaxies, which is likely an overestimation according to \cite{2005PASA...22..118G}. We do not see a significant reduction in the raw ICL+CG brightness when adjusting the masking apertures from 2.5 Kron radii to 3.5 Kron radii. Thus, we expect the "pure ICL"+CG profile computed in this paper to be an underestimation, and the truth "pure"  ICL+CG profile to be between the "pure" and "raw" measurements.

Finally, our cluster galaxy residual subtraction process does not consider the existence of the recently discovered ultra-diffuse galaxies \citep[see][and the referencing papers]{2015ApJ...798L..45V}. Neither have these galaxies been given consideration in the cluster galaxy luminosity function models used in the paper. Therefore, our "pure" ICL+CG profile contains light from these ultradiffuse galaxies. Distinguishing them will require a better understanding of these lesser-known objects in the future.

\section{Summary and Discussion}
\label{sec:conclude}

This paper develops methods to measure low surface brightness ICL centered on the cluster CG, which are tested with random points and simulated ICL+CG profiles, and also applied to the study of the DECam PSFs in the $g-$ and $r-$bands for averaged DES Y3 observations. We estimate the light residual in raw measurements of ICL+CG profiles from cluster satellite galaxies and consider systematic effects from the sky background subtraction procedure, from the extended wings of the DECam PSFs and from assumptions about the luminosity function of cluster satellite galaxies.

Following the high S/N detection of ICL+CG radial profiles, we study a variety of ICL properties -- radial distribution, integrated luminosity, color profile, and the connection between ICL distribution and the cluster mass distribution. The ICL extends out to 1 Mpc from the cluster centers at a surface brightness level of 30 mag arcsec$^{-2}$. The ICL outside 30 kpc of the CGs, is $\sim3$ times as luminous as the ICL+CG central component within 30 kpc. The ICL also displays an interesting self-similarity feature wherein their radial profiles appear similar after scaling by the cluster $R_{200m}$, and the ICL radial distribution closely follows the overall cluster mass distribution.

\subsection{Comparison to Previous Observational Studies}
\label{sec:compare_obs}

\begin{figure}
\includegraphics[width=0.45\textwidth]{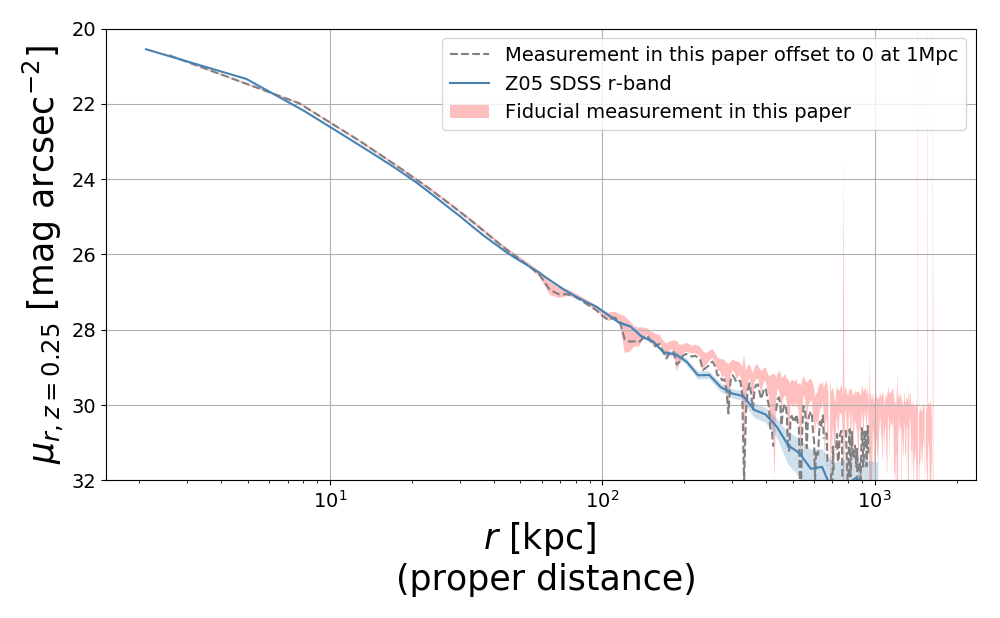}
\caption{Our ICL+CG radial profile (DES $r$-band, red shaded region) in comparison to the measurements in Z05 (SDSS $r$-band, blue shaded region). Our measurements agree remarkably well with Z05 within 100 kpc, yet are much brighter than Z05 outside 100 kpc. The different background estimation methods partially explain the differences -- when offsetting our ICL flux measurements to 0 at 1 Mpc (as is done in Z05), our measurement agrees better with Z05 (dashed gray line). See section~\ref{sec:compare_obs} for discussions.}
\label{fig:z05}
\end{figure}

A very similar work to this one is described in \cite{2005MNRAS.358..949Z} (hereafter Z05) wherein $g$, $r$, and $i$ images from the Sloan Digital Sky Survey (SDSS) of $\sim 600$ galaxy clusters at redshift 0.2-0.3 were stacked, and ICL+CG profiles out to 700 kpc from the CG centers were derived. Our analyses shared some similar ICL findings to Z05 --both works have found the ICL to have a shallower profile than the CGs in the core, bluer ICL $g-r$ color at a greater distance, and more luminous ICL in richer clusters, yet the core component of the CGs have a similar luminosity regardless of the cluster richness. 

Not all results in the two works agree, however. Notably, the ICL profile derived with our method is much brighter than that in Z05, as shown in Figure~\ref{fig:z05}. While Z05 estimated the ICL surface brightness to be 32 mag arcsec$^{-2}$ in the $r-$band at around 700 kpc, our results show that the ICL surface brightness is above 32 mag/arcsec$^{-2}$ even at 1 Mpc. Several methodological differences may have contributed to this discrepancy. Most importantly, Z05 chose to evaluate the sky background level as the average flux in a 100 kpc annulus centered at 1 Mpc from the cluster center, which offsets the ICL flux level to 0 in that radial range. Our nominal method offsets the ICL flux level to 0 at $\sim$ 1.8 Mpc, and if we offset our measured ICL flux to 0 in the same radial range as Z05, our ICL surface brightness drops to $\sim$ 31.5 mag arcsec$^{-2}$ at 700 kpc, close to the estimation in Z05. The rest of the differences may be due to the applications of K- and e-corrections (wavelength shifting with redshift and galaxy passive evolution), as well as cluster selection methods.  Overall, our analysis is based on a much improved cluster sample with better techniques to quantify sky background and PSFs and account for sky selection effects associated with the cluster sample.

Many other studies have investigated ICL properties to a great distance, albeit with fewer clusters  \cite[e.g., ][]{2006AJ....131..168K, 2007ApJ...666..147G, 2007AJ....134..466K, 2010ApJ...720..569R, 2012A&A...537A..64G, 2014ApJ...781...24G, 2014ApJ...794..137M,2015MNRAS.449.2353B}. With deep images from {\it Hubble} or ground-based  telescopes, these works have measured the ICL to surface brightness levels beyond 25 mag arcsec$^{-2}$ and out to at least 100 kpc from the CG centers, sometimes close to a surface brightness limit of 30 mag arcsec$^{-2}$. 

A frequently investigated ICL property in these works is the fractional contribution of the ICL to the total cluster stellar content. Around redshift 0.2-0.3, there is a great dispersion in the reported values. \cite{2015MNRAS.449.2353B} found the ICL fraction to be around 20\% at redshift 0.25 in the CLASH survey data, while \cite{2007AJ....134..466K} estimated the fraction to be 6\%-20\% in 10 nearby clusters, or around 10\% at redshift 0.25. Going to a slightly lower redshift range of 0.16-0.20, \cite{2004ApJ...615..196F} estimated the ICL fraction to be around 10\% with four Abell clusters. Note that these three works have defined the ICL as the diffuse light fainter than a surface brightness limit, usually around 25-26 mag arcsec$^{-2}$ in $B-$ or $V-$bands, which roughly correspond to $>$ 20 kpc in our analysis. On the higher side, \cite{2007ApJ...666..147G, 2013ApJ...778...14G} estimated the ICL+CG to make up 33\% of the total cluster stellar mass within $r_{200}$ in the redshift range of 0.03-0.13, or 20\%-50\% in the redshift range of 0.05-0.24 within $r_{500}$. \cite{2011MNRAS.414..602T} found the ICL+CG fraction to be $>$40\% within 500 kpc for one cluster at redshift 0.29, and \cite{2017ApJ...846..139M} found the fraction within 300 kpc to be between 15\% and 60\%. The reported ICL fraction varies with the radial range of the comparison. For one cluster of redshift 0.44, \cite{2014A&A...565A.126P} estimated the ICL+CG fraction to be $>$ 50\% within 100 $\mathrm{kpc/h}$, or 20 \% within 350 $\mathrm{kpc/h}$, or 8.2 \%within $R_{500}$. Our measurement of the ICL+CG luminosity fraction to be $44\pm17$\% at radius 1 Mpc is on the high end compared to the previous observations, although still within the range of those reports.

Another frequently visited ICL property is the color, which informs us about its stellar population. Most imaging studies have found ICL+CG colors to be comparable to an old stellar population with a radial trend of becoming bluer at a greater distance \citep[e.g.,][]{2014ApJ...794..137M, 2014A&A...565A.126P,  2017ApJ...834...16M, 2018MNRAS.475.3348H, 2018MNRAS.474.3009D}, which is further supported by spectroscopic studies of the ICL+CG stellar composition \citep{2016A&A...592A...7A, 2016MNRAS.461..230E, 2018arXiv180506913J}. Our results of the ICL color being consistent with the cluster red sequence galaxies and becoming bluer with distance agree with those previous conclusions (e.g., Z05, as shown in Figure~\ref{fig:color}). The color gradient of ICL in particular indicates that the ICL is likely produced from disruptions of dwarf galaxies or tidal stripping of cluster member galaxies. The former mechanism is expected to create an ICL color gradient as the disrupted dwarfs at different radii have different masses and colors. The later mechanism, the tidal stripping of galaxies, becomes stronger closer to the cluster center and will be able to strip the redder, inner parts of the cluster galaxies. Major mergers are not likely to be the main ICL formation mechanism, as it would give rise to a more uniform ICL color profile (e.g. \citealt{2013A&A...553A..99E, 2018arXiv181103253C}). 

\subsection{Comparison to Simulations}

A few simulation studies have attempted to include the ICL \citep{2006ApJ...648..936R, 2007MNRAS.377....2M,  2007ApJ...666...20P, 2011MNRAS.413..101G, 2011ApJ...732...48R} but balancing between the properties of the general cluster galaxy population and the properties of CG and ICL is a challenge \citep{2010MNRAS.406..936P}. 

Using semianalytical methods,  \cite{2014MNRAS.437.3787C, 2018MNRAS.479..932C} investigated different pathways to ICL formation, such as stripping and merging relaxation, and found that regardless of these mechanisms, the ICL has experienced rapid redshift evolution in the late times. At redshift 0, the ICL+CG may make up 30\%-50\% of the total cluster stellar content, or close to such a quantity at redshift 0.25. 

Recent developments in hydrodynamics simulations are also producing encouraging ICL results. In high-resolution dark matter and baryon resimulations of massive cluster-sized halos incorporating AGN feedback, \cite{2014MNRAS.440.2290M}  were able to identify diffuse ICL envelopes, extending to a few hundreds of kpcs around the CGs. Between surface brightness limits of 25 and 27 mag arcsec$^{-2}$ in $V-$band, ICL makes up 20\%-60\% of the total light of the CG and ICL combination. The surface brightness limits applied in \cite{2014MNRAS.440.2290M} roughly correspond to a 30-80 kpc radial range.  Our estimation of the fraction of the ICL in the ICL+CG combination in such a radial range is $28\%$ (Figure~\ref{fig:cluster_profile3}), which agrees with the reports in  \cite{2014MNRAS.440.2290M}.

 \cite{2018MNRAS.475..648P} carried out a detailed study of the ICL with the IllustrisTNG hydrodynamics simulation suites. In particular, \cite{2018MNRAS.475..648P} investigated the ICL radial distribution, and found the ICL stellar mass profile to be as shallow as that of the dark matter, with a similar power-law radial dependence.  The ICL stellar mass outside 30 kpc, or outside 100 kpc of the cluster center, scales with cluster mass with a power-law index close to 1. Our observation that ICL traces the overall cluster mass distribution supports such a finding. As with the amount of the ICL, for a $3\times10^{14} M_\odot$ cluster, \cite{2018MNRAS.475..648P}  found that the ICL stellar mass outside 30 kpc of the CG center is about three time as massive as the inner 30 kpc of the CG+ICL combination ($3\times10^{12} M_\odot$ versus $1\times10^{12} M_\odot$), and the ICL outside 100 kpc is approximately equal to the CG+ICL stellar mass within 100 kpc ($2\times10^{12} M_\odot$). The ICL+CG fraction in the total cluster stellar content is 50 \%, which is on the high end of the previous observational studies. However, both the \cite{2018MNRAS.475..648P} and the \cite{2014MNRAS.437.3787C, 2018MNRAS.479..932C} estimations are in excellent agreement with the results from our analysis.

Modeling-wise, \cite{2013ApJ...770...57B,2018arXiv180607893B} also included an ICL component, called intra-halo light (IHL), in their efforts to provide a self-consistent halo model to quantify the average galaxy star formation history from various galaxy observables. 
For a halo of mass $1\times10^{14} \mathrm{M_\odot}$, \cite{2018arXiv180607893B} also found that ICL/IHL has a rather late formation time, increasing rapidly by $\sim 3$ times between redshift 0 and 1. Near redshift 0,  the ICL/IHL contributes $\sim30$ \% of the total halo stellar mass.

\subsection{ICL Self-similarity}

A rather interesting result from this analysis is that the ICL radial profile appears self-similar -- it scales with cluster $R_{200m}$ inferred from cluster richness. In addition,  starting from 100 kpc and out to 1 Mpc, the ICL profile appears to trace a theoretical cluster mass profile model. This is one of the first direct evidence establishing a connection between the ICL and the cluster mass radial distribution. Prior to this work, \cite{2018arXiv180711488M} examined the surface brightness contours of the ICL and the weak lensing mass maps of six galaxy clusters studied by the {\it Hubble} frontier program and found (1) visual similarity between the two and moreover, (2) compatible contour shapes with a quantitative shape estimator, in various radial ranges. \cite{2018arXiv180711488M} noted the potential of using the ICL observation to trace the cluster mass distribution. The results from our analysis provide additional direct evidence to such a conclusion.

Prior to the analyses in this paper and those presented in \cite{2018arXiv180711488M},  hints of a connection between the ICL and cluster mass distributions can be found in a few observational and simulation studies that examine the radial profile of the ICL or the scaling between the ICL stellar mass and cluster mass. For example, \cite{2005MNRAS.358..949Z} found that the ICL profile is reasonably approximated by an NFW model \citep{1997ApJ...490..493N}. A few other observational works have noted a stronger correlation between cluster mass and ICL stellar mass or luminosity outside the CG cores \citep[e.g., ][]{2018MNRAS.474.3009D, 2018MNRAS.480..521H}. While the accuracy of the ICL tracing cluster mass distribution awaits further investigations, this similarity may provide another channel to explore the origin of the ICL, if different ICL formation mechanisms, e.g., tidal stripping or dwarf galaxy disruption, have different radial or cluster mass dependence. Furthermore, as noted in \cite{2018arXiv180711488M}, this phenomenon, if confirmed, provides another extraordinary opportunity to observationally constrain the cluster mass distribution. 

Another possible explanation of this phenomenon is that the ICL simply traces the luminosity distribution of cluster satellite galaxies, which then traces the cluster mass distribution. We have examined the radial light profile of the cluster satellite galaxies derived in Section ~\ref{sec:pure_icl} and compared it to the theoretical cluster mass profile model, but cannot determine if the ICL is a better or worse tracer of cluster mass than the satellite galaxies, given the large uncertainties of those measurements. It would be worthwhile to revisit this topic with a larger cluster sample in the future.

\subsection{Outlook}

With the successful detection ofthe ICL in this paper, an interesting topic to explore next is its redshift evolution, which will provide more clues about its origin.  The depth and wavelength coverage of the DES produce volume-limited cluster samples up to redshift $\sim 0.65$ in Y1 data and up to redshift $\sim 0.8$ with the first three years of data (Y3).  The increased volume of the \rdmp cluster sample at a higher redshift and hence the increased sample size partially offset the dimming effect of the ICL with distance and may allow for ICL detection up to redshift $\sim 0.65$, which will provide preliminary answers to whether or not the ICL has experienced much redshift evolution.

The method developed in this paper can also be applied to studying low surface brightness light around other types of galaxies in DES data. We note that we have employed this method to analyze the faint light halo around luminous red galaxies \citep{2016MNRAS.461.1431R} in an upcoming study (Leung  in preparation).

Finally, based on the ICL surface brightness results presented in this work, the ICL may become a significant systematic effect for cluster weak lensing studies with future cosmic surveys such as the LSST, because of ICL contamination in the flux measurements of weak lensing source galaxies. For this aspect of the ICL study, we refer the readers to \cite{2018arXiv180904599G}.

\acknowledgements
Some of the ancillary data products shown in the Figures of this paper are available on the DES data release page
\url{https://des.ncsa.illinois.edu/releases/other/paper-data}. Readers interested in comparing to these results are encouraged to check out the release page or contact the corresponding author for additional information. 

We thank the anonymous referee for the constructive comments and Stefano Zibetti for kindly sharing the data products in his paper for comparison with this work.
We also thank Rebecca Bernstein, Gary Bernstein, Peter Behroozi, Risa Weschler, Tom McClintock and Eduardo Rozo for insightful discussions. We use Monte Carlo Markov Chain sampling in this analysis, which is performed with the PYMC \citep{Patil10pymc:bayesian} and emcee \citep{2013PASP..125..306F} Python packages.

Funding for the DES Projects has been provided by the U.S. Department of Energy, the U.S. National Science Foundation, the Ministry of Science and Education of Spain, 
the Science and Technology Facilities Council of the United Kingdom, the Higher Education Funding Council for England, the National Center for Supercomputing 
Applications at the University of Illinois at Urbana-Champaign, the Kavli Institute of Cosmological Physics at the University of Chicago, 
the Center for Cosmology and Astro-Particle Physics at the Ohio State University,
the Mitchell Institute for Fundamental Physics and Astronomy at Texas A\&M University, Financiadora de Estudos e Projetos, 
Funda{\c c}{\~a}o Carlos Chagas Filho de Amparo {\`a} Pesquisa do Estado do Rio de Janeiro, Conselho Nacional de Desenvolvimento Cient{\'i}fico e Tecnol{\'o}gico and 
the Minist{\'e}rio da Ci{\^e}ncia, Tecnologia e Inova{\c c}{\~a}o, the Deutsche Forschungsgemeinschaft and the Collaborating Institutions in the Dark Energy Survey. 

The Collaborating Institutions are Argonne National Laboratory, the University of California at Santa Cruz, the University of Cambridge, Centro de Investigaciones Energ{\'e}ticas, 
Medioambientales y Tecnol{\'o}gicas-Madrid, the University of Chicago, University College London, the DES-Brazil Consortium, the University of Edinburgh, 
the Eidgen{\"o}ssische Technische Hochschule (ETH) Z{\"u}rich, 
Fermi National Accelerator Laboratory, the University of Illinois at Urbana-Champaign, the Institut de Ci{\`e}ncies de l'Espai (IEEC/CSIC), 
the Institut de F{\'i}sica d'Altes Energies, Lawrence Berkeley National Laboratory, the Ludwig-Maximilians Universit{\"a}t M{\"u}nchen and the associated Excellence Cluster Universe, 
the University of Michigan, the National Optical Astronomy Observatory, the University of Nottingham, The Ohio State University, the University of Pennsylvania, the University of Portsmouth, 
SLAC National Accelerator Laboratory, Stanford University, the University of Sussex, Texas A\&M University, and the OzDES Membership Consortium.

Based in part on observations at Cerro Tololo Inter-American Observatory, National Optical Astronomy Observatory, which is operated by the Association of 
Universities for Research in Astronomy (AURA) under a cooperative agreement with the National Science Foundation.

The DES data management system is supported by the National Science Foundation under Grant Numbers AST-1138766 and AST-1536171.
The DES participants from Spanish institutions are partially supported by MINECO under grants AYA2015-71825, ESP2015-66861, FPA2015-68048, SEV-2016-0588, SEV-2016-0597, and MDM-2015-0509, 
some of which include ERDF funds from the European Union. IFAE is partially funded by the CERCA program of the Generalitat de Catalunya.
Research leading to these results has received funding from the European Research
Council under the European Union's Seventh Framework Program (FP7/2007-2013) including ERC grant agreements 240672, 291329, and 306478.
We  acknowledge support from the Australian Research Council Centre of Excellence for All-sky Astrophysics (CAASTRO), through project number CE110001020, and the Brazilian Instituto Nacional de Ci\^encia
e Tecnologia (INCT) e-Universe (CNPq grant 465376/2014-2).

This manuscript has been authored by Fermi Research Alliance, LLC under Contract No. DE-AC02-07CH11359 with the U.S. Department of Energy, Office of Science, Office of High Energy Physics. The United States Government retains and the publisher, by accepting the article for publication, acknowledges that the United States Government retains a non-exclusive, paid-up, irrevocable, world-wide license to publish or reproduce the published form of this manuscript, or allow others to do so, for United States Government purposes.

\appendix
\section{Examples of Noise Sources in ICL Detection}

\begin{figure*}
\includegraphics[width=1.0\textwidth]{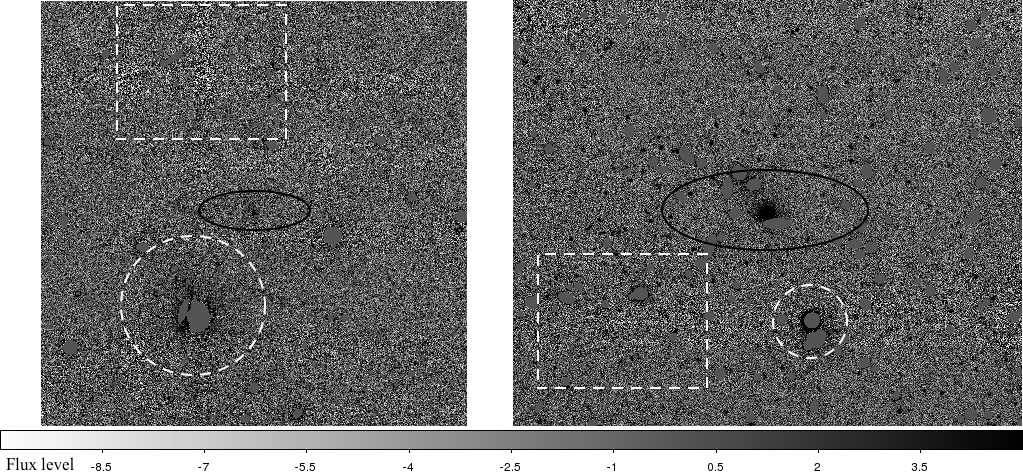}
\caption{Masked $r$ band images of one cluster in our analysis, centered on the CG. The left panel shows the whole 4000$\times$4000 pixels (1 pixel = 0.263$\arcsec$) image, and the right panel is a DS9 1/2 zoom-in of the same cluster. Bright galaxies or stars above 22.4 magnitude in $i$ band other than the CG have been masked, but the unmasked fainter objects can be seen in the right panel. The black ellipticals indicate the CG location. The dotted circles indicate regions with bright foreground stars which show visible unmasked light residue. The dotted boxes indicate regions with sky flux discontinuities from coadding multiple single exposures. The gray bar below the image indicates the flux level while both of the images have a magnitude zeropoint of 30. }
\label{fig:singlestack}
\end{figure*}

\begin{figure*}
\includegraphics[width=1.0\textwidth]{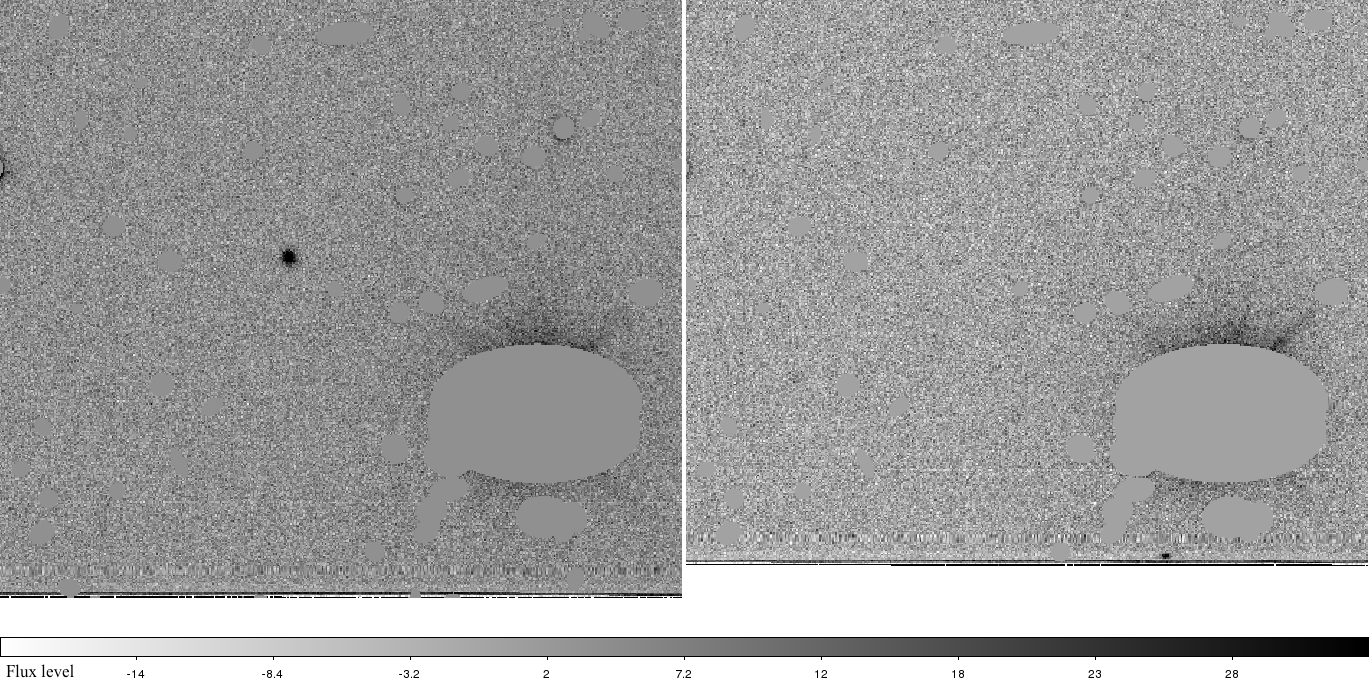}
\caption{Masked $g-$ band images of one cluster in our analysis which shows an unmasked transient in one of the exposures. The right image was taken on 2013 September 9, and the left image was taken on 2013 November 22. The transient disappeared in the left image. This object does not appear in the DES coadd object catalog, and hence is not masked in our image-processing steps. We have masked all the other objects detected in the DES coadd object catalog to produce these figures. 
}
\label{fig:transient}
\end{figure*}

While examining the masked images of the clusters, we identified a few cases that increase the noise level of ICL detection. Three of them are noted below as examples. These noted contaminations are not structures associated with the redshift 0.2-0.3 clusters, therefore we expect their net effects to be canceled out in the process of "stacking" and background subtraction. However, these cases can contribute correlated noise on a scale of a few pixels, or produce the noise spikes seen in the ICL figures.

\begin{itemize}
\item Bright stars or nearby galaxies (e.g., galaxies below redshift 0.05). The unmasked light residual in the outskirts of these objects can be bright enough to cause significant contamination to the faint ICL detection on a scale of tens of arcseconds. Examples of such cases can be seen in Figure~\ref{fig:singlestack}. Note that we have implemented a random point subtraction process to account for the fact that the cluster sample selection method avoids looking for clusters in close proximity to these objects.
\item Coadding discontinuity in the images. This can be caused by different background residual levels in the processed single-exposure images, or CCD readout differences in the same exposure. Because DES does not purposely place clusters at certain locations in the camera field of view, this effect should be canceled out in the "stacking" process. An example of such a case can be seen in Figure~\ref{fig:singlestack}.
\item Transient objects (e.g., astroids). We noticed these objects as they may appear to be particularly bright in one exposure but disappear in another. These objects are often not cataloged in the DES coadd object database, as the object-detection process is only based on the coadded images of DES $r-$, $i-$ and $z-$band observations. For the same reason, these contamination cases are more common in the $g-$band rather than in the $r-$band. Again, their contamination can be canceled out by "stacking". An example of such a case is shown in Figure~\ref{fig:transient}.
\end{itemize}

\bibliography{reference}

\end{document}

%% file: icl_authors.tex

\def\andname{}

\author{
Y.~Zhang\altaffilmark{1,$\dagger$},
B.~Yanny\altaffilmark{1},
A.~Palmese\altaffilmark{1},
D.~Gruen\altaffilmark{2,3},
C.~To\altaffilmark{4},
E.~S.~Rykoff\altaffilmark{3,2},
Y.~Leung\altaffilmark{5},
C.~Collins\altaffilmark{6},
M.~Hilton\altaffilmark{7},
T.~M.~C.~Abbott\altaffilmark{8},
J.~Annis\altaffilmark{1},
S.~Avila\altaffilmark{9},
E.~Bertin\altaffilmark{10,11},
D.~Brooks\altaffilmark{12},
D.~L.~Burke\altaffilmark{2,3},
A.~Carnero~Rosell\altaffilmark{13,14},
M.~Carrasco~Kind\altaffilmark{15,16},
J.~Carretero\altaffilmark{17},
C.~E.~Cunha\altaffilmark{2},
C.~B.~D'Andrea\altaffilmark{18},
L.~N.~da Costa\altaffilmark{14,19},
J.~De~Vicente\altaffilmark{13},
S.~Desai\altaffilmark{20},
H.~T.~Diehl\altaffilmark{1},
J.~P.~Dietrich\altaffilmark{21,22},
P.~Doel\altaffilmark{12},
A.~Drlica-Wagner\altaffilmark{1,5},
T.~F.~Eifler\altaffilmark{23,24},
A.~E.~Evrard\altaffilmark{25,26},
B.~Flaugher\altaffilmark{1},
P.~Fosalba\altaffilmark{27,28},
J.~Frieman\altaffilmark{1,5},
J.~Garc\'ia-Bellido\altaffilmark{29},
E.~Gaztanaga\altaffilmark{27,28},
D.~W.~Gerdes\altaffilmark{25,26},
R.~A.~Gruendl\altaffilmark{15,16},
J.~Gschwend\altaffilmark{14,19},
G.~Gutierrez\altaffilmark{1},
W.~G.~Hartley\altaffilmark{12,30},
D.~L.~Hollowood\altaffilmark{31},
K.~Honscheid\altaffilmark{32,33},
B.~Hoyle\altaffilmark{34,35},
D.~J.~James\altaffilmark{36},
T.~Jeltema\altaffilmark{31},
K.~Kuehn\altaffilmark{37},
N.~Kuropatkin\altaffilmark{1},
T.~S.~Li\altaffilmark{1,5},
M.~Lima\altaffilmark{38,14},
M.~A.~G.~Maia\altaffilmark{14,19},
M.~March\altaffilmark{18},
J.~L.~Marshall\altaffilmark{39},
P.~Melchior\altaffilmark{40},
F.~Menanteau\altaffilmark{15,16},
C.~J.~Miller\altaffilmark{25,26},
R.~Miquel\altaffilmark{41,17},
J.~J.~Mohr\altaffilmark{21,22,34},
R.~L.~C.~Ogando\altaffilmark{14,19},
A.~A.~Plazas\altaffilmark{24},
A.~K.~Romer\altaffilmark{42},
E.~Sanchez\altaffilmark{13},
V.~Scarpine\altaffilmark{1},
M.~Schubnell\altaffilmark{26},
S.~Serrano\altaffilmark{27,28},
I.~Sevilla-Noarbe\altaffilmark{13},
M.~Smith\altaffilmark{43},
M.~Soares-Santos\altaffilmark{44},
F.~Sobreira\altaffilmark{45,14},
E.~Suchyta\altaffilmark{46},
M.~E.~C.~Swanson\altaffilmark{16},
G.~Tarle\altaffilmark{26},
D.~Thomas\altaffilmark{9},
W.~Wester\altaffilmark{1}
\\ \vspace{0.2cm} (DES Collaboration) \\
}

\affil{$^{1}$ Fermi National Accelerator Laboratory, P. O. Box 500, Batavia, IL 60510, USA}
\affil{$^{2}$ Kavli Institute for Particle Astrophysics \& Cosmology, P. O. Box 2450, Stanford University, Stanford, CA 94305, USA}
\affil{$^{3}$ SLAC National Accelerator Laboratory, Menlo Park, CA 94025, USA}
\affil{$^{4}$ Department of Physics, Stanford University, 382 Via Pueblo Mall, Stanford, CA 94305, USA}
\affil{$^{5}$ Kavli Institute for Cosmological Physics, University of Chicago, Chicago, IL 60637, USA}
\affil{$^{6}$ Astrophysics Research Institute, Liverpool John Moores University, IC2 Building, Liverpool Science Park, 146 Brownlow Hill, L3 5RF, UK}
\affil{$^{7}$ Astrophysics \& Cosmology Research Unit, School of Mathematics, Statistics \& Computer Science, University of KwaZulu-Natal, Westville Campus, Private Bag X54001, Durban 4000, South Africa}
\affil{$^{8}$ Cerro Tololo Inter-American Observatory, National Optical Astronomy Observatory, Casilla 603, La Serena, Chile}
\affil{$^{9}$ Institute of Cosmology and Gravitation, University of Portsmouth, Portsmouth, PO1 3FX, UK}
\affil{$^{10}$ CNRS, UMR 7095, Institut d'Astrophysique de Paris, F-75014, Paris, France}
\affil{$^{11}$ Sorbonne Universit\'es, UPMC Univ Paris 06, UMR 7095, Institut d'Astrophysique de Paris, F-75014, Paris, France}
\affil{$^{12}$ Department of Physics \& Astronomy, University College London, Gower Street, London, WC1E 6BT, UK}
\affil{$^{13}$ Centro de Investigaciones Energ\'eticas, Medioambientales y Tecnol\'ogicas (CIEMAT), Madrid, Spain}
\affil{$^{14}$ Laborat\'orio Interinstitucional de e-Astronomia - LIneA, Rua Gal. Jos\'e Cristino 77, Rio de Janeiro, RJ - 20921-400, Brazil}
\affil{$^{15}$ Department of Astronomy, University of Illinois at Urbana-Champaign, 1002 W. Green Street, Urbana, IL 61801, USA}
\affil{$^{16}$ National Center for Supercomputing Applications, 1205 West Clark St., Urbana, IL 61801, USA}
\affil{$^{17}$ Institut de F\'{\i}sica d'Altes Energies (IFAE), The Barcelona Institute of Science and Technology, Campus UAB, 08193 Bellaterra (Barcelona) Spain}
\affil{$^{18}$ Department of Physics and Astronomy, University of Pennsylvania, Philadelphia, PA 19104, USA}
\affil{$^{19}$ Observat\'orio Nacional, Rua Gal. Jos\'e Cristino 77, Rio de Janeiro, RJ - 20921-400, Brazil}
\affil{$^{20}$ Department of Physics, IIT Hyderabad, Kandi, Telangana 502285, India}
\affil{$^{21}$ Excellence Cluster Universe, Boltzmannstr.\ 2, 85748 Garching, Germany}
\affil{$^{22}$ Faculty of Physics, Ludwig-Maximilians-Universit\"at, Scheinerstr. 1, 81679 Munich, Germany}
\affil{$^{23}$ Department of Astronomy/Steward Observatory, 933 North Cherry Avenue, Tucson, AZ 85721-0065, USA}
\affil{$^{24}$ Jet Propulsion Laboratory, California Institute of Technology, 4800 Oak Grove Dr., Pasadena, CA 91109, USA}
\affil{$^{25}$ Department of Astronomy, University of Michigan, Ann Arbor, MI 48109, USA}
\affil{$^{26}$ Department of Physics, University of Michigan, Ann Arbor, MI 48109, USA}
\affil{$^{27}$ Institut d'Estudis Espacials de Catalunya (IEEC), 08034 Barcelona, Spain}
\affil{$^{28}$ Institute of Space Sciences (ICE, CSIC),  Campus UAB, Carrer de Can Magrans, s/n,  08193 Barcelona, Spain}
\affil{$^{29}$ Instituto de Fisica Teorica UAM/CSIC, Universidad Autonoma de Madrid, 28049 Madrid, Spain}
\affil{$^{30}$ Department of Physics, ETH Zurich, Wolfgang-Pauli-Strasse 16, CH-8093 Zurich, Switzerland}
\affil{$^{31}$ Santa Cruz Institute for Particle Physics, Santa Cruz, CA 95064, USA}
\affil{$^{32}$ Center for Cosmology and Astro-Particle Physics, The Ohio State University, Columbus, OH 43210, USA}
\affil{$^{33}$ Department of Physics, The Ohio State University, Columbus, OH 43210, USA}
\affil{$^{34}$ Max Planck Institute for Extraterrestrial Physics, Giessenbachstrasse, 85748 Garching, Germany}
\affil{$^{35}$ Universit\"ats-Sternwarte, Fakult\"at f\"ur Physik, Ludwig-Maximilians Universit\"at M\"unchen, Scheinerstr. 1, 81679 M\"unchen, Germany}
\affil{$^{36}$ Harvard-Smithsonian Center for Astrophysics, Cambridge, MA 02138, USA}
\affil{$^{37}$ Australian Astronomical Optics, Macquarie University, North Ryde, NSW 2113, Australia}
\affil{$^{38}$ Departamento de F\'isica Matem\'atica, Instituto de F\'isica, Universidade de S\~ao Paulo, CP 66318, S\~ao Paulo, SP, 05314-970, Brazil}
\affil{$^{39}$ George P. and Cynthia Woods Mitchell Institute for Fundamental Physics and Astronomy, and Department of Physics and Astronomy, Texas A\&M University, College Station, TX 77843,  USA}
\affil{$^{40}$ Department of Astrophysical Sciences, Princeton University, Peyton Hall, Princeton, NJ 08544, USA}
\affil{$^{41}$ Instituci\'o Catalana de Recerca i Estudis Avan\c{c}ats, E-08010 Barcelona, Spain}
\affil{$^{42}$ Department of Physics and Astronomy, Pevensey Building, University of Sussex, Brighton, BN1 9QH, UK}
\affil{$^{43}$ School of Physics and Astronomy, University of Southampton,  Southampton, SO17 1BJ, UK}
\affil{$^{44}$ Brandeis University, Physics Department, 415 South Street, Waltham MA 02453}
\affil{$^{45}$ Instituto de F\'isica Gleb Wataghin, Universidade Estadual de Campinas, 13083-859, Campinas, SP, Brazil}
\affil{$^{46}$ Computer Science and Mathematics Division, Oak Ridge National Laboratory, Oak Ridge, TN 37831}